\documentclass[sigconf,hyphens,spaces,obeyspaces,acmsmall]{acmart} %acmart
\usepackage{url}
\usepackage{tabularx}
\usepackage{color}
\usepackage{etoolbox}
\usepackage{listings}
\usepackage{xcolor}
\usepackage{multirow}
\usepackage{pifont}
\usepackage{enumitem}
\usepackage{subfig}
\urldef{\footurl}\url{https://midiaresearch.com/blog/announcing-midias-state-of-the-streaming-nation-2-report}

\colorlet{punct}{red!60!black}
\definecolor{background}{HTML}{EEEEEE}
\definecolor{delim}{RGB}{20,105,176}
\colorlet{numb}{magenta!60!black}
\makeatletter
\patchcmd{\maketitle}{\@copyrightspace}{}{}{}
\makeatother

\makeatletter 
\makeatletter

\newcommand{\cmark}{\ding{51}}%
\newcommand{\xmark}{\ding{55}}%

\acmConference[ACM Transactions on Intelligent Systems and Technology]{ACM}{September}{2018}
\newif\ifworkinprogress
 \workinprogresstrue

\ifworkinprogress
	\newcommand{\ms}[1]{\textcolor{blue}{\textbf{[Markus] #1}}}
	\newcommand{\pl}[1]{\textcolor{magenta}{\textbf{[Paul] #1}}}
	\newcommand{\hz}[1]{\textcolor{cyan}{\textbf{[Hamed] #1}}}
	\newcommand{\cw}[1]{\textcolor{green}{\textbf{[Ching-Wei] #1}}}
\else
  \newcommand{\ms}[1]{}
  \newcommand{\pl}[1]{}
  \newcommand{\cw}[1]{}
  \newcommand{\hz}[1]{}
\fi

\lstdefinelanguage{json}{
    basicstyle=\normalfont\ttfamily,
    float=tp,
    numbers=left,
    numberstyle=\scriptsize,
    stepnumber=1,
    numbersep=8pt,
    showstringspaces=false,
    breaklines=true,
    frame=lines,
    backgroundcolor=\color{background},
    literate=
      {:}{{{\color{punct}{:}}}}{1}
      {,}{{{\color{punct}{,}}}}{1}
      {\{}{{{\color{delim}{\{}}}}{1}
      {\}}{{{\color{delim}{\}}}}}{1}
      {[}{{{\color{delim}{[}}}}{1}
      {]}{{{\color{delim}{]}}}}{1},
}

\begin{document}

\title{Report on ACM Recommender Systems Challenge 2018: Automatic Music Playlist Continuation}
\title{ACM RecSys Challenge 2018: Automatic Music Playlist Continuation (Final Report)}
\title{The ACM RecSys Challenge 2018 on Automatic Music Playlist Continuation}
\title{An Analysis of Approaches Taken in the ACM RecSys Challenge 2018 for Automatic Music Playlist Continuation}

%\renewcommand{\shortauthors}{}
%\renewcommand{\shorttitle}{}

\author{Hamed Zamani}
\affiliation{
  \institution{University of Massachusetts Amherst} % Center for Intelligent Information Retrieval, 
  \city{Amherst} 
  \state{USA}
}
\email{zamani@cs.umass.edu}

\author{Markus Schedl}
\affiliation{
  \institution{Johannes Kepler University Linz}  % Department of Computational Perception, 
  \city{Linz} 
  \state{Austria} 
}
\email{markus.schedl@jku.at}

\author{Paul Lamere}
\affiliation{
  \institution{Spotify}
  \city{New York} 
  \state{USA} 
}
\email{paul@spotify.com}

\author{Ching-Wei Chen}
\affiliation{
  \institution{Spotify}
  \city{New York} 
  \state{USA} 
}
\email{cw@spotify.com}

\begin{abstract}
The ACM Recommender Systems Challenge 2018 focused on the task of automatic music playlist continuation, which is a form of the more general task of sequential recommendation. Given a playlist of arbitrary length with some additional meta-data, the task was to recommend up to 500 tracks that fit the target characteristics of the original playlist. 
For the RecSys Challenge, Spotify released a dataset of one million user-generated playlists.
Participants could compete in two tracks, i.e., main and creative tracks. Participants in the main track were only allowed to use the provided training set, however, in the creative track, the use of external public sources was permitted.
In total, 113 teams submitted 1,228 runs to the main track; 33 teams submitted 239 runs to the creative track.
The highest performing team in the main track achieved an R-precision of $0.2241$, an NDCG of $0.3946$, and an average number of recommended songs clicks of $1.784$.
In the creative track, an R-precision of $0.2233$, an NDCG of $0.3939$, and a click rate of $1.785$ was obtained by the best team. This article provides an overview of the challenge, including motivation, task definition, dataset description, and evaluation. We further report and analyze the results obtained by the top performing teams in each track and explore the approaches taken by the winners. We finally summarize our key findings, discuss generalizability of approaches and results to domains other than music, and list the open avenues and possible future directions in the area of automatic playlist continuation.
\end{abstract}




% \setcopyright{acmlicensed}
% \acmJournal{TIST}
% \acmYear{2019} \acmVolume{1} \acmNumber{1} \acmArticle{1} \acmMonth{1} \acmPrice{15.00}\acmDOI{10.1145/3344257}

\maketitle

\section{Overview}
\label{sec:overview}
According to a study carried out in 2016 by the Music Business Association\footnote{\url{https://musicbiz.org/news/playlists-overtake-albums-listenership-says-loop-study}} as part of their Music Biz Consumer Insights program,\footnote{\url{https://musicbiz.org/resources/tools/music-biz-consumer-insights/consumer-insights-portal}} 
playlists accounted for 31\% of music listening time among listeners in the United States, which is more than albums (22\%), but less than single tracks (46\%). 
In a 2017 study conducted by Nielsen,\footnote{\url{http://nielsen.com/us/en/insights/reports/2017/music-360-2017-highlights.html}} it was found that 58\% of users in the United States create their own playlists, 32\% share them with others. 
Other studies, conducted by MIDiA,\footnote{\footurl} 
show that 55\% of music streaming service subscribers create music playlists, using streaming services. 
Studies like these suggest a growing importance of playlists as a mode of music consumption, which is also reflected in the fact that the music streaming service Spotify currently hosts over 2 billion playlists.\footnote{\url{https://press.spotify.com/us/about}}

In its most generic definition, a playlist is simply a sequence of tracks intended to be listened to together. The task of automatic playlist generation then refers to the automated creation of these sequences of tracks~\cite{bonnin2015}. In this context, the ordering of songs\footnote{In this paper, the terms ``song'' and ``track'' are used, interchangeably.} in a playlist is often highlighted as a key characteristics of automatic playlist generation, which makes the task a highly complex endeavor. Some authors have therefore proposed approaches based on Markov chains to model the transitions between songs in playlists, e.g.~\cite{chen_etal:kdd:2012,mcfee_lanckriet:ismir:2011}. While these approaches have been shown to outperform approaches agnostic of the song order in terms of log likelihood, recent research has found little evidence that the exact order of songs actually matters to users~\cite{Tintarev:2017:SDS:3079628.3079633}, while the ensemble of songs in a playlist~\cite{vall_etal:recsys:2017} and direct song-to-song transitions~\cite{kamehkhosh_etal:milc:2018} seems to matter.

Considered a variation of automatic playlist generation, the task of \textit{automatic playlist continuation} (APC) consists of adding one or more tracks to a playlist in a way that fits the same target characteristics of the original playlist~\cite{Schedl2018,bonnin2015}. This has benefits in both the listening and creation of playlists: users can enjoy listening to continuous sessions beyond the end of a finite-length playlist, while also finding it easier to create longer, more compelling playlists without a need to have extensive musical familiarity.

Schedl et al. \cite{Schedl2018} have recently identified the task of automatic music playlist continuation as one of the grand challenges in music recommender systems research. A large part of the APC task is to accurately infer the intended purpose of a given playlist. This is challenging not only because of the broad range of these intended purposes (when they even exist), but also because of the diversity in the underlying features or characteristics that might be needed to infer those purposes. 

An extreme cold start scenario for this task is where a playlist is created with some meta-data (e.g., the title of a playlist), but no song has been added to the playlist. This problem can be cast as an \emph{ad-hoc information retrieval task}, where the task is to rank songs in response to a user-provided meta-data query. 

Given the importance of playlists in improving the user experience within the context of music streaming services, ACM Recommender Systems Challenge\footnote{ACM Recommender Systems Challenge, or RecSys Challenge in short, is an annual competition organized in conjunction with the ACM Conference on Recommender Systems, since 2010. For more information, refer to \cite{Said:2016} or visit \url{http://recsyschallenge.com/}.} 2018 \cite{Chen:2018} has focused on an automatic music playlist continuation task.\footnote{\url{http://2018.recsyschallenge.com}} This paper provides an overview of the challenge, the results achieved by over 100 participating teams as well as the winning and most innovative approaches, and future directions and open avenues in this research area.

\subsection{Task: Automatic Playlist Continuation}
As mentioned earlier, automatic playlist continuation is a useful feature for music streaming services not only because it can extend listening session length, but also because it can increase engagement of users on their platform by making it easier for users to create playlists that they can enjoy and share. ACM Recommender Systems Challenge 2018 has focused on the task of automatic playlist continuation (APC).
This task consists of adding one or more tracks to a music playlist in a way that fits the target characteristics of the original playlist~\cite{Schedl2018,bonnin2015}. To formally define the task, let $\mathcal{M}$ be the universe of tracks in the underlying music catalog. Given a playlist $P$ created by a user $u$, that contains $k$ music tracks $M_P = \{m_{P1}, m_{P2}, \cdots, m_{Pk}\}$, the task is to rank the music tracks from $\mathcal{M}-M_P$ to be recommended to the user for completing the playlist. In addition, each playlist includes some meta-data information, such as title. It should be noted that $k$ can be equal to zero for some playlists, meaning that the user has created the playlist but no music track has yet been added to the playlist. 

\subsection{Competition: Main and Creative Tracks}
ACM Recommender Systems Challenge 2018 invited participants to submit their solutions for the APC task in two distinct tracks: main track and creative track. Participants in the main track were only allowed to use the dataset provided by the challenge for training their models. In contrast, participants in the creative track were required to use external resources, such as public datasets, for solving the same task. The submitted solutions for both tracks were evaluated using the same dataset, which will be explained in the following subsection.

% \hz{main and creative tracks should be explained later. in results probably}
% \ms{i think it makes sense to do it already here. and briefly again later (with a pointer to the longer description herew).}
% Participants had to devise algorithms that predict, for a given playlist, an ordered list of 500 recommended candidate tracks. Performance was evaluated against a challenge set (cf.~Section~\ref{sec:dataset}) of user-created playlists, where different combinations of playlist titles and some numbers of tracks were withheld. 
% Their algorithms could either use only the data in the provided training dataset or may additionally exploit publicly available external data sources. Submissions provided by algorithms of the former kind were considered for the main track, those of the latter kind for the creative track. 

\subsection{Data: Million Playlist Dataset}
\label{sec:data}
For algorithm development and testing, we released a dataset of one million user-created playlists from the Spotify platform, dubbed the \textit{Million Playlist Dataset} (MPD). These playlists were created during the period of January 2010 until November 2017. Statistics of the MPD are reported in Table~\ref{tab:mpd_stats}. The dataset includes, for each playlist, its title as well as the list of tracks (including album and artist names), and some additional meta-data such as Spotify URIs and the playlist's number of followers.  
The playlist titles in the dataset were unmodified, however for reporting in Table~\ref{tab:mpd_stats}, playlist titles were lightly normalized by converting to lowercase, and removing spaces and common non-alphanumeric symbols. A truncated sample playlist is shown in Appendix \ref{appendix:sample_playlist}. 

\begin{table}[t!]
\centering
\caption{Basic statistics of the Million Playlist Dataset.}
\begin{tabular}{lr}\hline\hline
Property & Value \\\hline
Number of playlists & 1,000,000\\
Number of tracks & 66,346,428 \\
Number of unique tracks & 2,262,292 \\
Number of unique albums & 734,684 \\
Number of unique artists & 295,860 \\
Number of unique playlist titles & 92,944 \\
Number of unique normalized playlist titles & 17,381 \\
Average playlist length (tracks) & 66.35 \\
\hline\hline
\end{tabular}
\label{tab:mpd_stats}
\end{table}

A separate \textit{challenge dataset} was used to validate the quality of the elaborated algorithms. It consisted of a set of playlists from which a number of tracks had been withheld. 
The challenge set was composed of 10,000 incomplete playlists and covered a total of 10 scenarios (1000 playlists for each): (1) title only, no track, (2) title and the first 5 tracks, (3) the first 5 tracks, (4) title and the first 10 tracks, (5) the first 10 tracks, (6) title and the first 25 tracks, (7) title and 25 random tracks, (8) title and the first 100 tracks, (9) title and 100 random tracks, and (10) title and the first track. 

The task was then to predict the missing tracks in those playlists, and participating teams were required to submit their predictions for those missing tracks (as a list of 500 ordered predictions). 
The withheld tracks were used by the organizers as ground truth, i.e.~to compute the performance measures for each submission.

Note that the data provided by the challenge does not contain acoustic information or features. However, participants in the creative track were able to use the Spotify API (or other sources) to retrieve such information.

In order to foster reproducibility and further research in music recommendation, the dataset will be made available for researchers on the Spotify Research website.\footnote{\url{https://research.spotify.com/datasets}}

\subsection{Evaluation}
\label{sec:eval}
To assess the quality of submissions, we computed three metrics and averaged them across all playlists in the challenge dataset: R-precision, normalized discounted cumulative gain (NDCG), and recommended songs clicks. The formal definition of these metrics is presented in Appendix~\ref{appendix:eval}.

The higher the R-precision and NDCG, the better. However, lower recommended songs clicks indicates better performance. To aggregate the individual scores for the three metrics, Borda rank aggregation~\cite{borda:1781} is used, i.e.~scores are converted to ranks, which are then summed up over the three measures to obtain a single performance score.

\section{Participation}
% \hz{todo: update with the info from Ching-Wei}

\begin{figure}[t]
    \centering
    \includegraphics[width=0.7\linewidth]{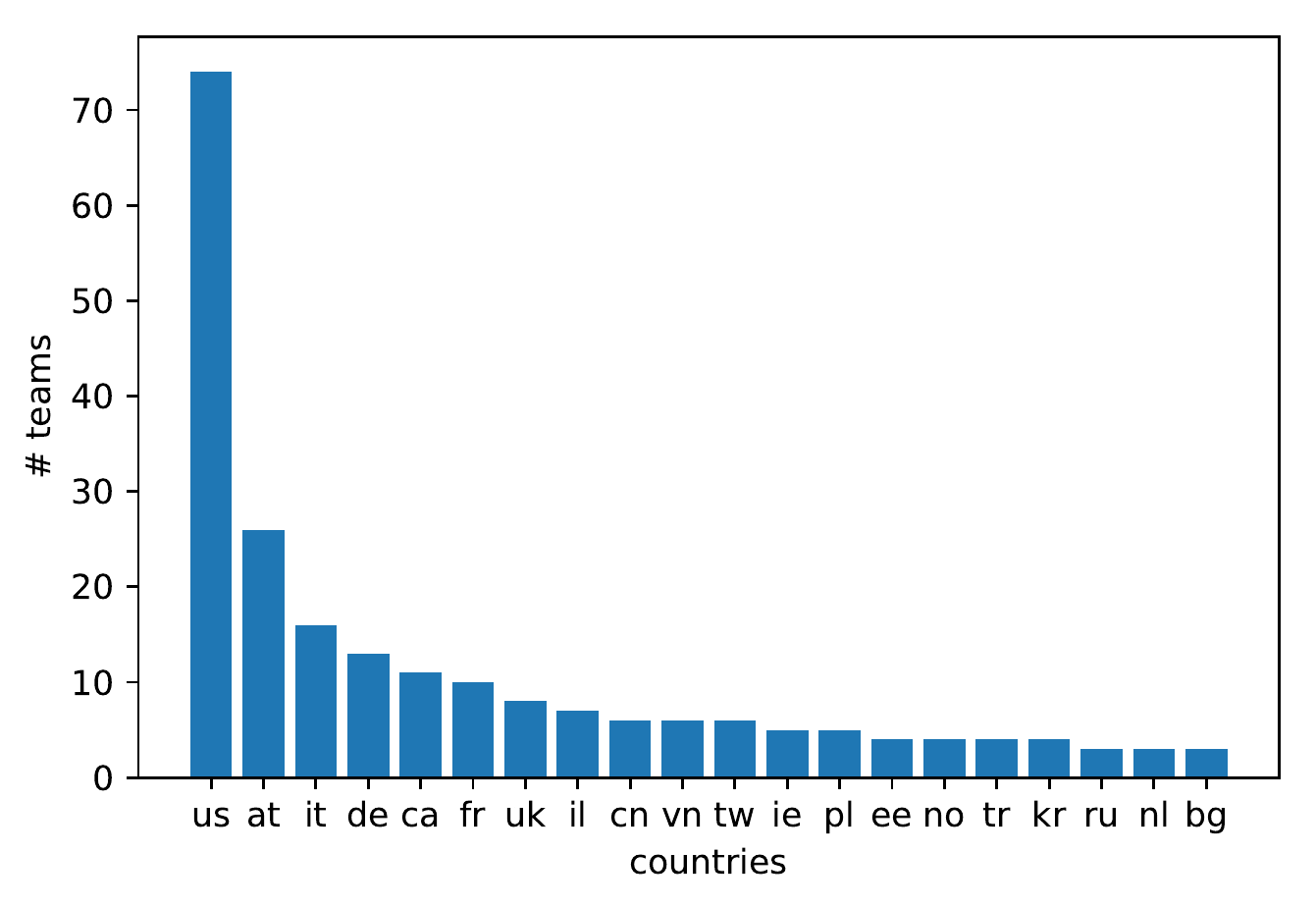}
    \caption{Number of registered teams per country for the top 20 countries.}
    \label{fig:n_teams_per_country}
\end{figure}

\begin{figure}[t]
    \centering
    \includegraphics[width=0.7\linewidth]{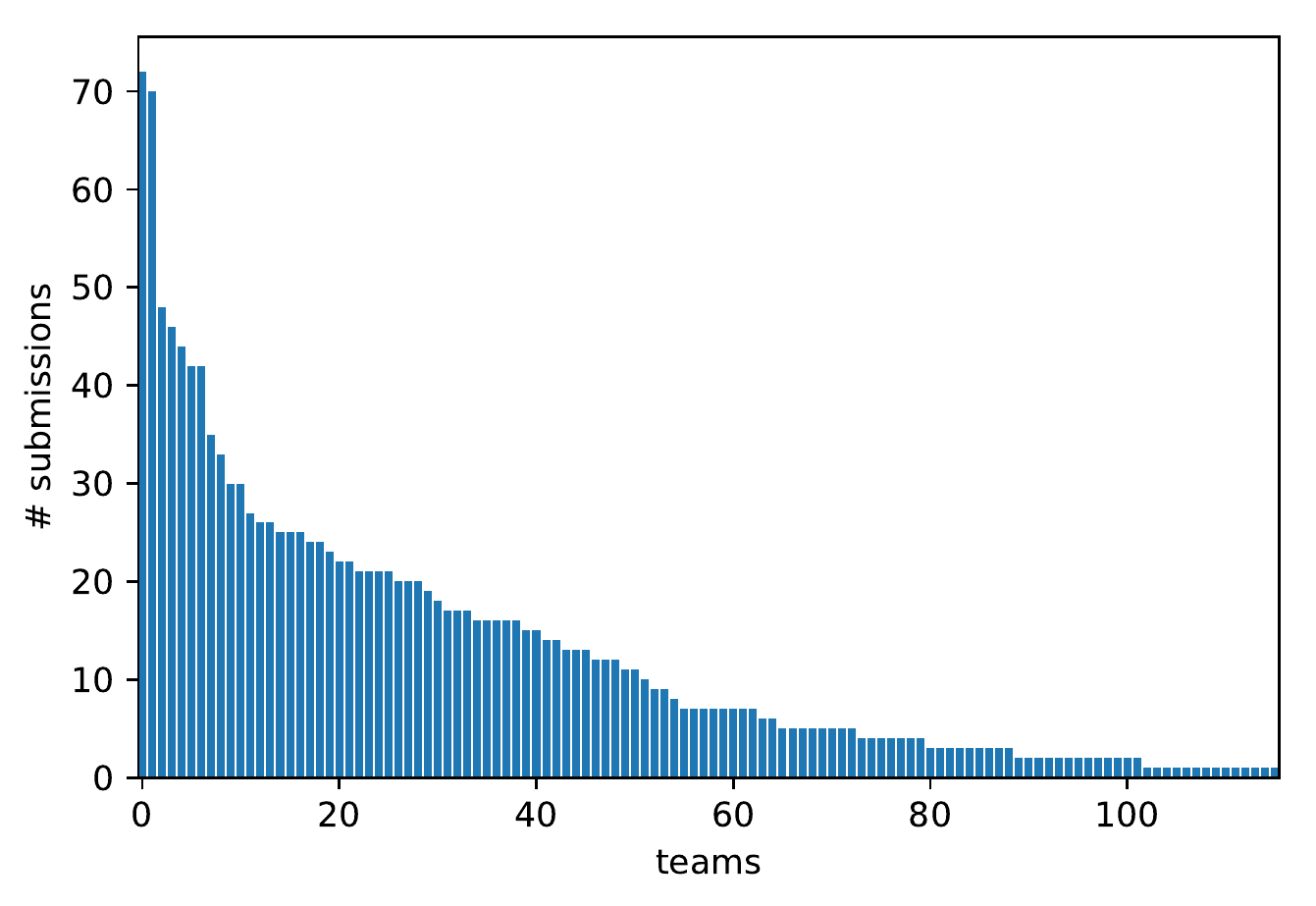}
    \caption{Number of submissions per team in descending order.}
    \label{fig:n_submissions_per_team}
\end{figure}

The RecSys Challenge was well received: 1,791 people registered; 1,430 with an academic affiliation and 361 from industry. These people formed a total of 410 teams. Out of these, 117 teams were active, i.e.,~submitted at least one run (113 and 33, respectively, to the main and to the creative track). The number of active teams per country for the top 20 countries (in terms of the number of teams) is plotted in Figure~\ref{fig:n_teams_per_country}. As depicted, the United States has the highest number of active teams followed by Austria and Italy. 

In total we received 1,467 submissions, out of which 1,228 were submitted to the main track and 239 to the creative track. The number of submissions made by each active team is plotted in Figure~\ref{fig:n_submissions_per_team}. 

% \hz{
% \begin{itemize}
%     \item \# teams per country
%     \item \# active teams (with at least one submission) per country
%     \item best result on the leader board over time for both tracks
%     \item the performance of vl6 over time for both tracks
%     \item \# of submitted runs per team
% \end{itemize}
% }

\section{Results}
The final results achieved by the participating teams for both main and creative tracks are available online.\footnote{The final leaderboard for the main track: \url{http://www.recsyschallenge.com/2018/leaderboard-main.html}}\textsuperscript{,}\footnote{The final leaderboard for the creative track: \url{http://www.recsyschallenge.com/2018/leaderboard-creative.html}} Tables~\ref{tab:main} and~\ref{tab:creative} summarize the results achieved by the top 10 teams in the main and creative tracks, respectively. Note that the test set for both tracks are the same and the only difference is that the teams were allowed to use external resources (other than the MPD training set) in the creative track.

As shown in Tables~\ref{tab:main} and~\ref{tab:creative}, the team vl6 has achieved the first ranked in both tracks, followed by teams hello word! and Avito in the main track and Creamy Fireflies and KAENEN in the creative track. The first ranked team has achieved the best results in terms of all evaluation metrics, except for the recommended songs clicks metric in the main track where it has been beaten by team Avito. 

\begin{figure*}[t]
\centering
\begin{minipage}{.33\linewidth}
\centering
\subfloat{\includegraphics[scale=.33]{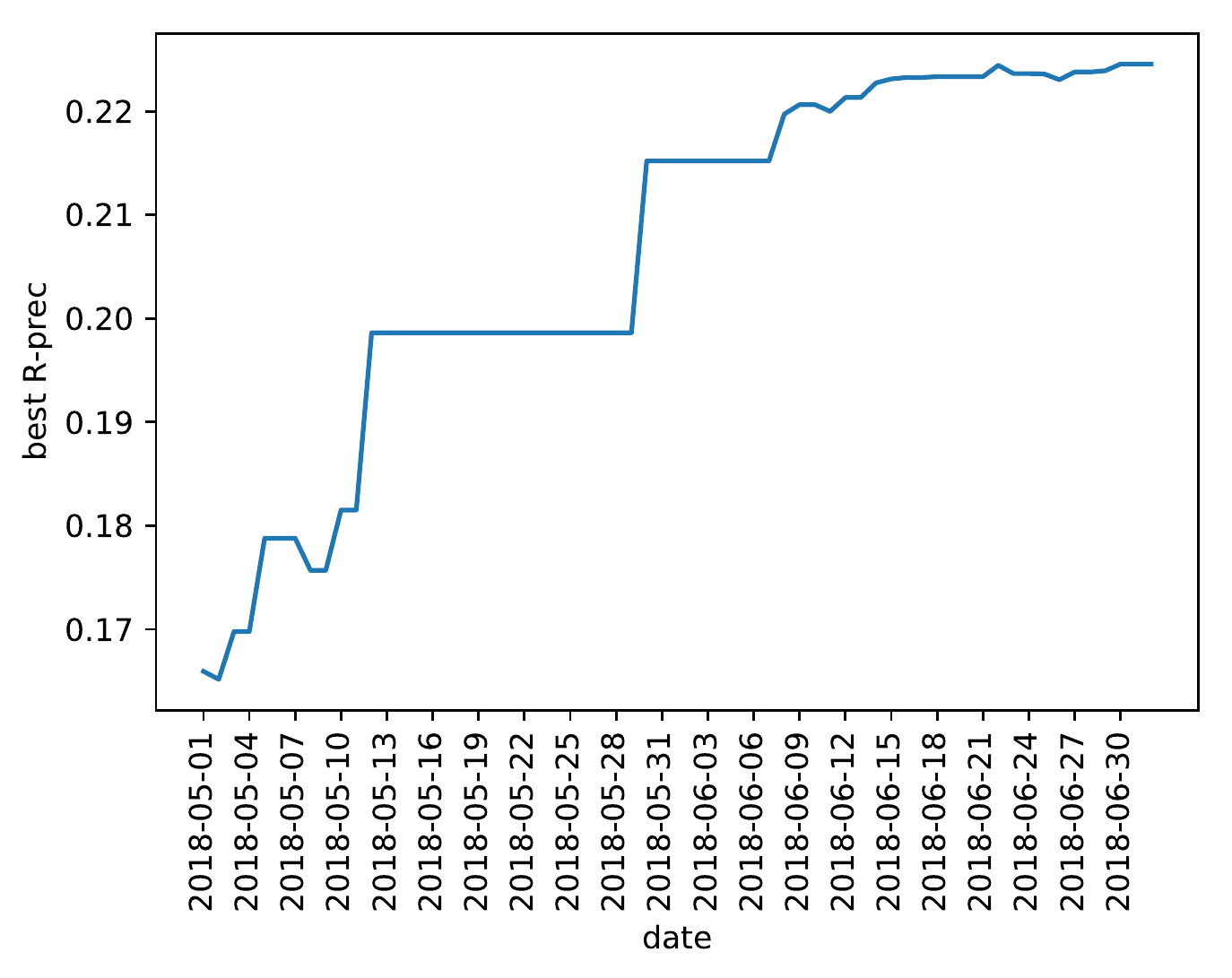}}
\end{minipage}%
\begin{minipage}{.33\linewidth}
\centering
\subfloat{\includegraphics[scale=.33]{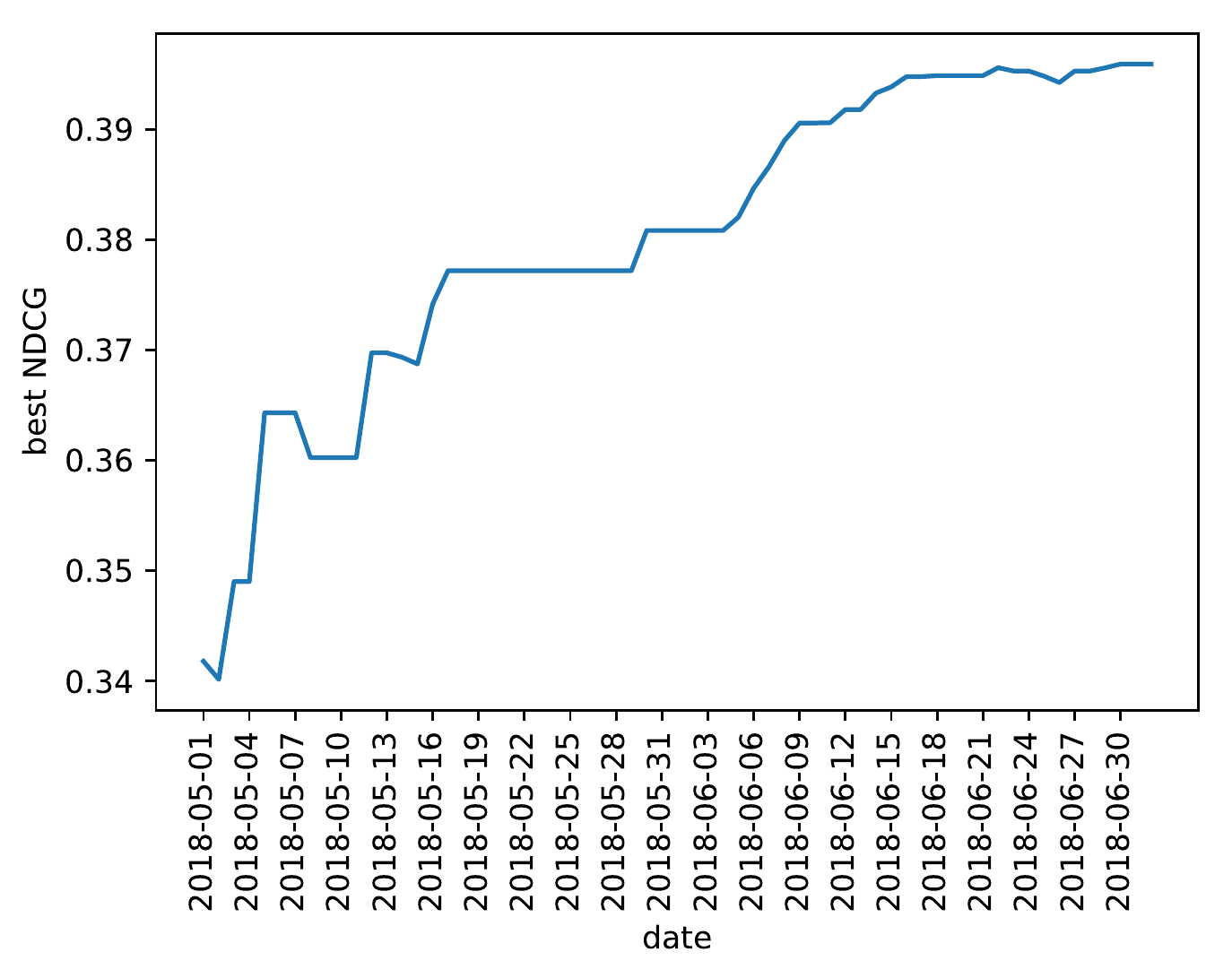}}
\end{minipage}
\begin{minipage}{.33\linewidth}
\centering
\subfloat{\includegraphics[scale=.33]{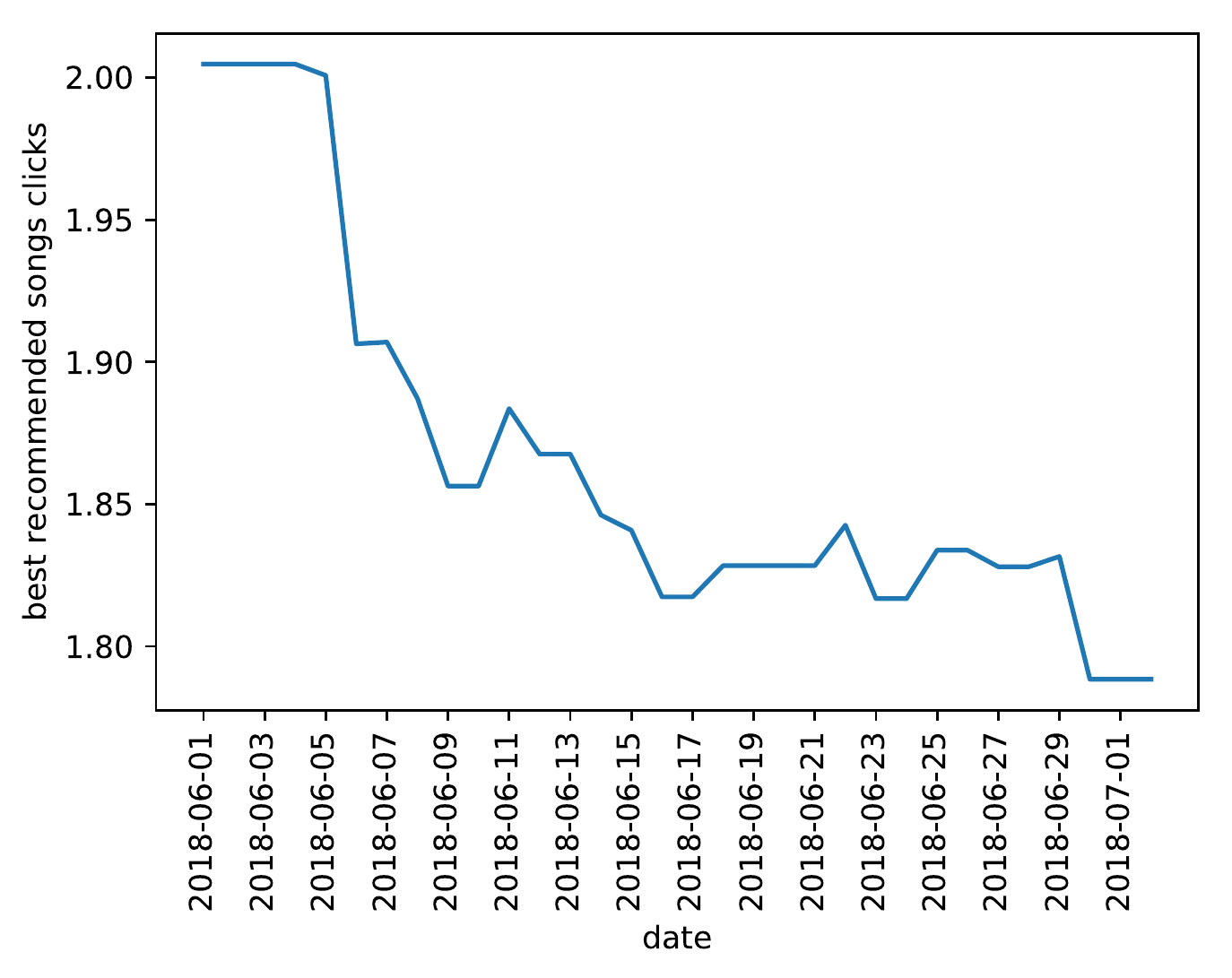}}
\end{minipage}%
\caption{The best performance in the leaderboard of the main track over time.} 
\label{fig:main-performance} 
\end{figure*}

\begin{figure*}[t]
\centering
\begin{minipage}{.33\linewidth}
\centering
\subfloat{\includegraphics[scale=.33]{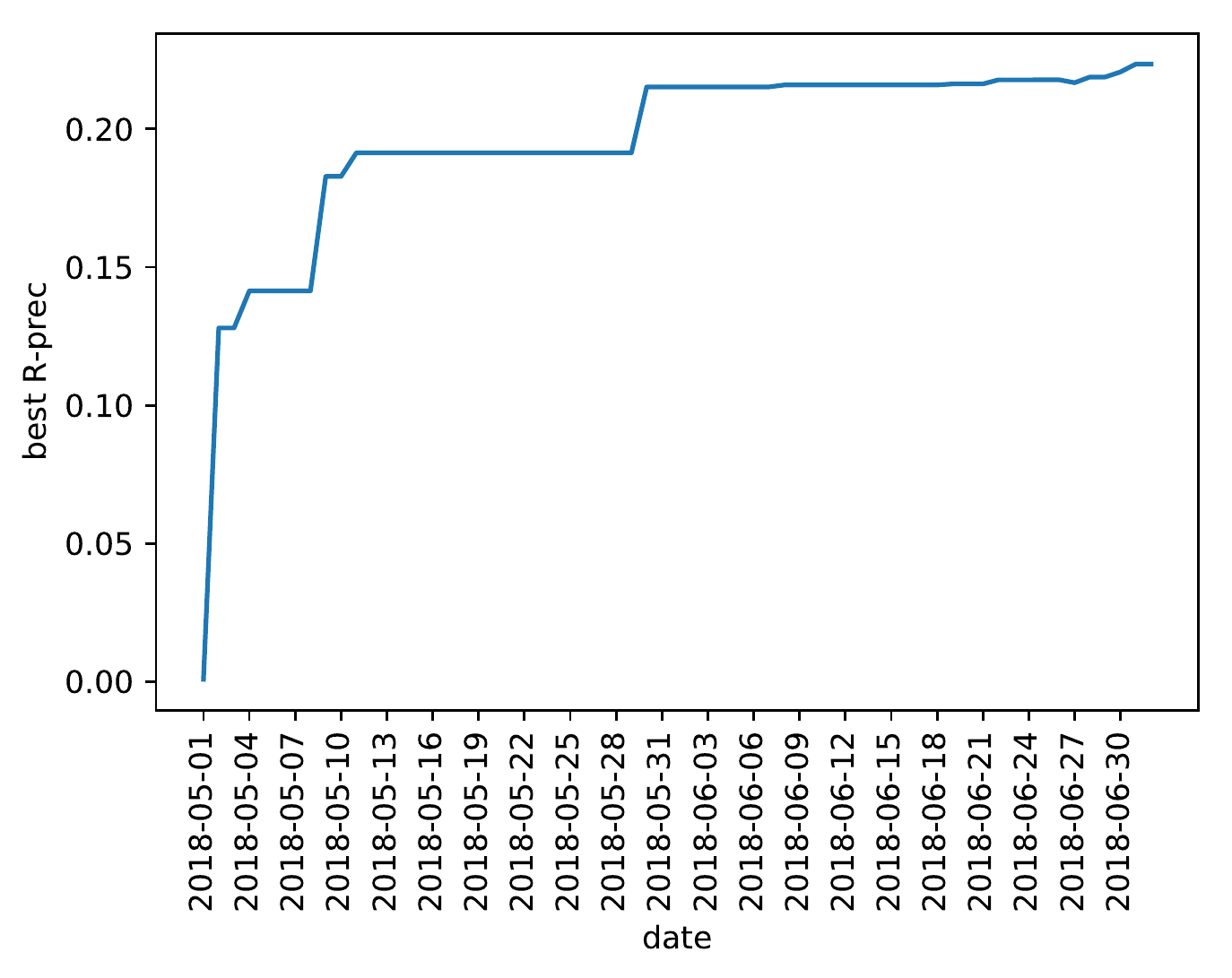}}
\end{minipage}%
\begin{minipage}{.33\linewidth}
\centering
\subfloat{\includegraphics[scale=.33]{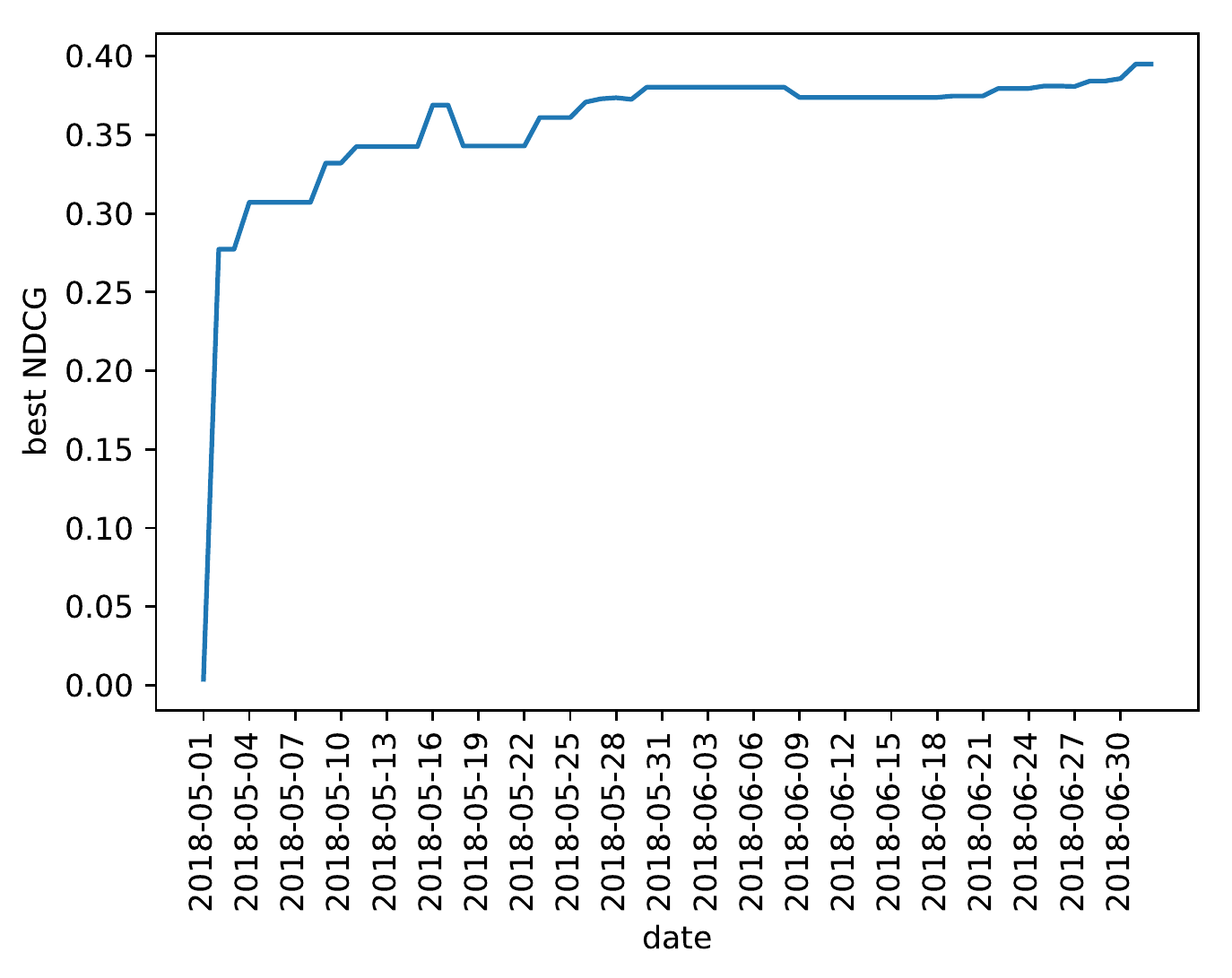}}
\end{minipage}
\begin{minipage}{.33\linewidth}
\centering
\subfloat{\includegraphics[scale=.33]{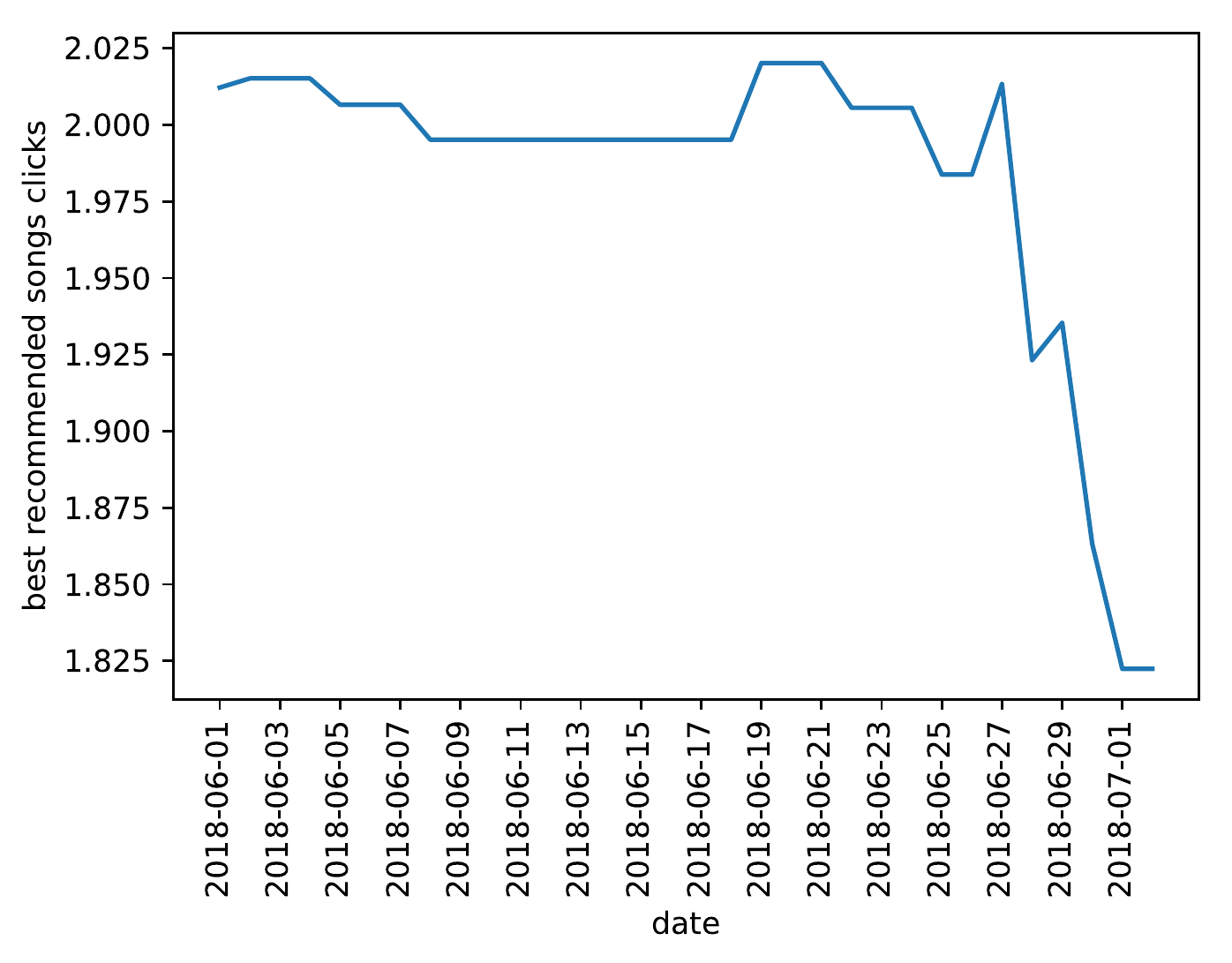}}
\end{minipage}%
\caption{The best performance in the leaderboard of the creative track over time.} 
\label{fig:creative-performance} 
\end{figure*}

\begin{figure*}[t]
\centering
\begin{minipage}{.33\linewidth}
\centering
\subfloat{\includegraphics[scale=.33]{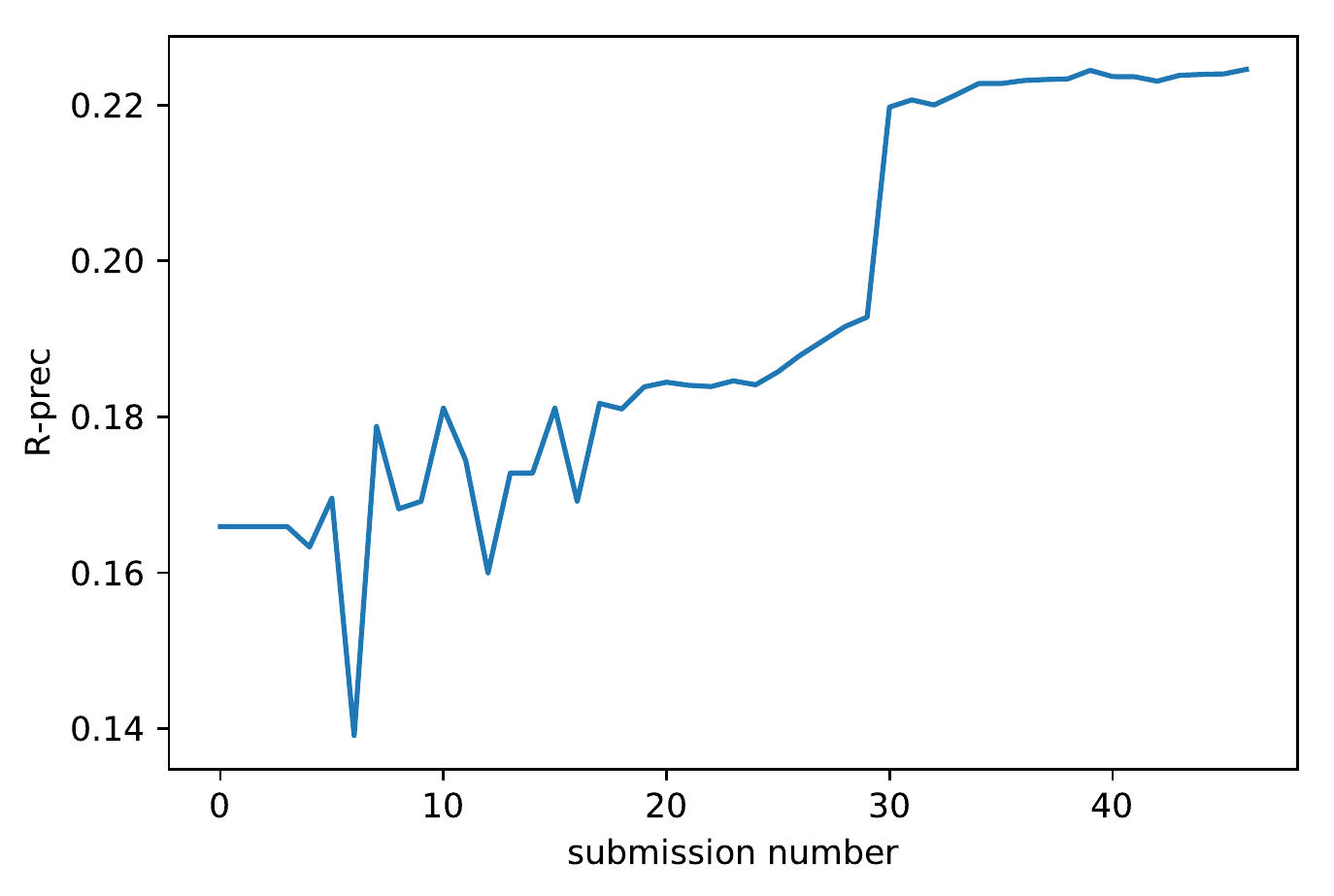}}
\end{minipage}%
\begin{minipage}{.33\linewidth}
\centering
\subfloat{\includegraphics[scale=.33]{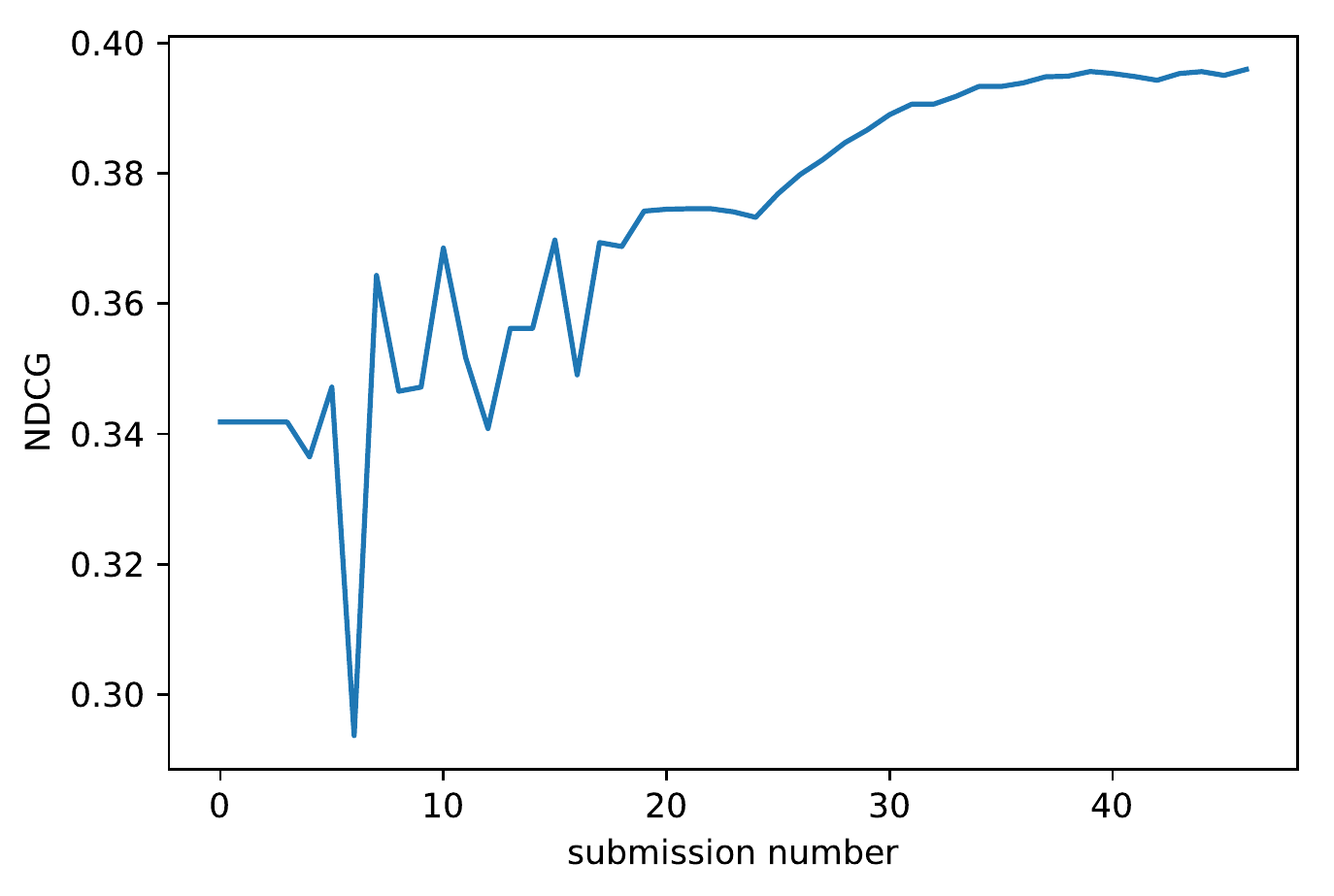}}
\end{minipage}
\begin{minipage}{.33\linewidth}
\centering
\subfloat{\includegraphics[scale=.33]{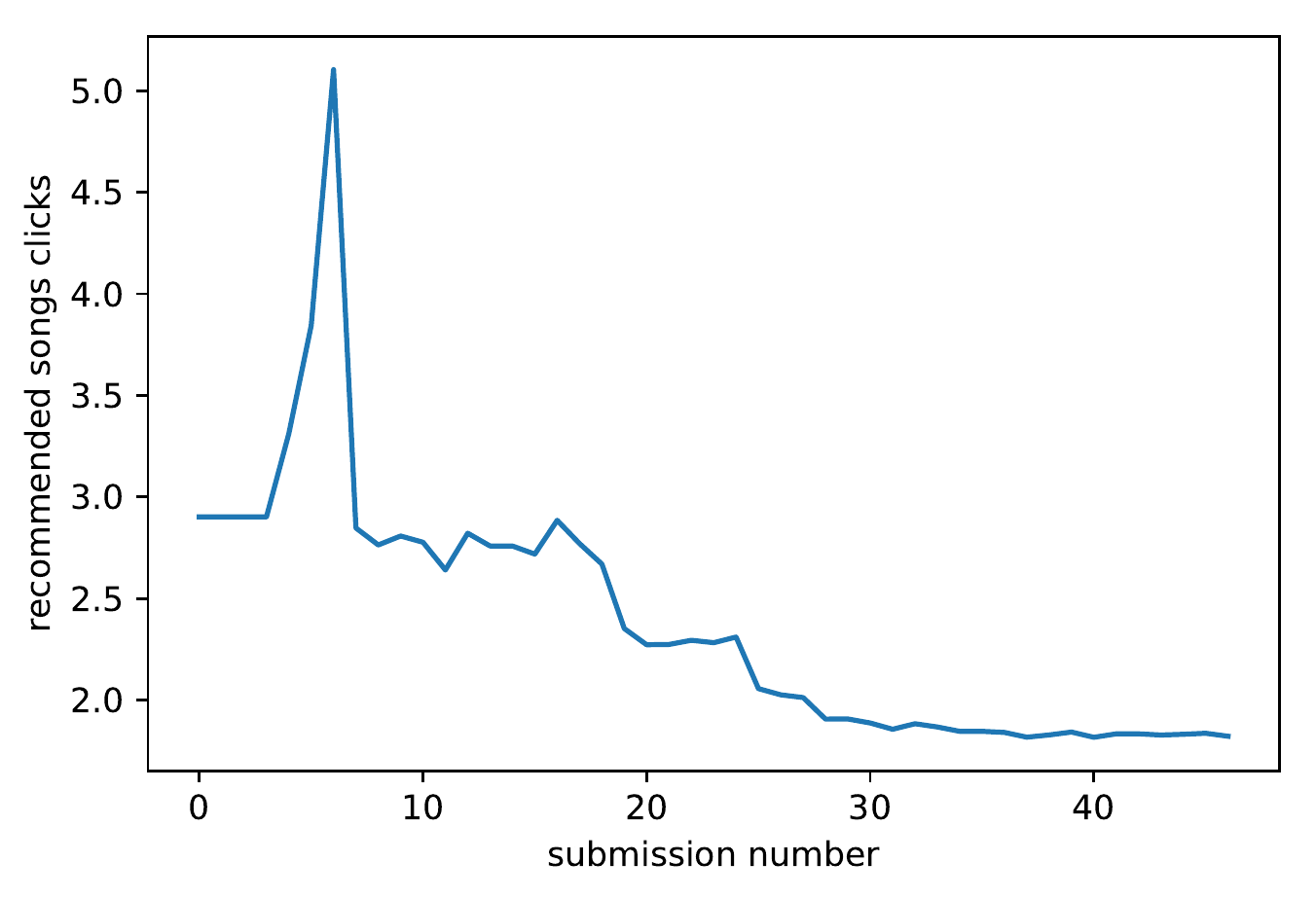}}
\end{minipage}%
\caption{The performance of the first ranked team (team vl6) in the main track over time.} 
\label{fig:vl6-performance} 
\end{figure*}

Figures~\ref{fig:main-performance} and \ref{fig:creative-performance} demonstrate the highest performance achieved in the leaderboard over time for the main and the creative tracks, respectively.\footnote{The starting date for the plots corresponding to recommended songs clicks differs from the starting dates in the other plots. This is due to the error of our evaluation script, which has been solved on 2018-06-01.} As expected, there is an increasing trend in terms of R-precision and NDCG and a decreasing trend in terms of recommended songs clicks over time. We also plot the performance of the first ranked team (team vl6) per submission over time in Figure~\ref{fig:vl6-performance}.

To gain a deep understanding of the performance of the models, we report the results for the 10 different types of playlists, separately (see Tables~\ref{tab:main:detailed} and \ref{tab:creative:detailed} for the main and creative tracks, respectively). As mentioned earlier in Section~\ref{sec:data}, the challenge set includes 10,000 playlists; 1000 playlists from each of the following playlist types: 1) title only, no track, (2) title and the first 5 tracks, (3) the first 5 tracks, (4) title and the first 10 tracks, (5) the first 10 tracks, (6) title and the first 25 tracks, (7) title and 25 random tracks, (8) title and the first 100 tracks, (9) title and 100 random tracks, and (10) title and the first track.

By analyzing the results reported in Tables~\ref{tab:main:detailed} and \ref{tab:creative:detailed}, we arrive at the following conclusions:

\begin{itemize}[leftmargin=*]
    \item As expected, by increasing the number of tracks as the input, the performance generally increases. There exist some exceptions, specially when 100 tracks are given. The reason can be due to the way that the teams handle the relation between the playlists. It is well known that most learning models fail at modeling long sequences, which also happens in the APC task.
    
    \item Surprisingly, the models perform worse when the title is also given as a meta-data for the playlist. For instance, the only difference between Type 2 and Type 3 is that the former contains playlist title. We believe that this strange behavior is observed because titles are highly sparse and models overfit on the titles appearing in the training set. In summary, the models fail at modeling the titles effectively.
    
    \item Interestingly, APC given random tracks produces much better results compared to the first tracks in the playlist (see the results for Type 6 vs. Type 7 and Type 8 vs. Type 9). This is due to the fact that adjacent tracks in a playlist are likely to share similar information, such at genre, artist, album, etc. Therefore, random tracks would provide more useful information to better understand the focus of the playlist, and thus more accurate APC performance is achieved.
    
    \item When the number of given tracks are more than or equal to 5, the recommended songs clicks for all the models is less than 1. This means that most users can find a relevant track in the top 10 recommended list and do not need to reload the recommended track list.
    
    \item By increasing the number of given tracks, the standard deviation of the performances obtained by the top 10 teams generally increases. In other words, most approaches perform closely when a few tracks are given. However, when several tracks are given for each playlist (e.g., more than or equal to 25 tracks), a substantial difference between the performance of different approaches is observed.
    
    \item Even one track matters: comparing the results of the playlists from Type 1 and Type 10, we observe a significant increase in the performance by adding only the first track of the playlist. This might be also due to the fact that the proposed solutions could not handle the title desirably. 
    
    \item In general, the team hello world! performed well when the first tracks of the playlists are given. However, the teams vl6 and MIPT\_MSU achieved the best results when the tracks are given in a random order. The team Avito also achieved the highest performance multiple times for some of the playlists that contain a few tracks.
    
    \item The performance of the models in the main track is slightly higher than that in the creative track. The reason might be that adding external resources increases the complexity of the models and given the amount of training data, the models could not take advantage of external resources, effectively. 
\end{itemize}

The approaches used by the top performing teams are briefly described in the next two sections.

\begin{table*}[t]
    \centering
    \caption{Final results achieved by the top 10 teams in the main track. The highest R-prec and NDCG as well as the lowest clicks are marked as bold.}
    \scalebox{0.75}{
    \begin{tabular}{llcccl}\hline\hline
        Rank & Team & R-prec & NDCG & clicks & code \\ \hline
        1 & vl6 \cite{Volkovs:vl6} & \textbf{0.2241} & \textbf{0.3946} & 1.7839 & \url{https://github.com/layer6ai-labs/vl6_recsys2018} \\
        2 & hello world! \cite{Yang:hello-world} & 0.2234 & 0.3932 & 1.8952 & \url{https://github.com/hojinYang/spotify_recSys_challenge_2018}\\
        3 & Avito \cite{Rubtsov:avito} & 0.2153 & 0.3846 & \textbf{1.7818} & \url{https://github.com/VasiliyRubtsov/recsys2018} \\
        4 & Creamy Fireflies \cite{Antenucci:creamy-fireflies} & 0.2202 & 0.3857 & 1.9335 & \url{https://github.com/tmscarla/spotify-recsys-challenge} \\
        4 & MIPT\_MSU & 0.2167 & 0.3823 & 1.8754 & \url{https://github.com/zakharovas/RecSys2018} \\
        6 & HAIR \cite{Zhu:hair}& 0.2163 & 0.3803 & 2.1815 & \url{https://github.com/LauraBowenHe/Recsys-Spotify-2018-challenge} \\
        7 & KAENEN \cite{Ludewig:kaenen}& 0.2091 & 0.3747 & 2.0540 & \url{https://github.com/rn5l/rsc18} \\
        7 & BachPropagate \cite{Kallumadi:bach-propagate} & 0.2090 & 0.3740 & 1.8834 & \url{https://bachpropagate.weebly.com/} \\
        9 & Definitive Turtles \cite{Kelen:definitive-turtles} & 0.2086 & 0.3751 & 2.0781 & \url{https://github.com/proto-n/recsys-challenge-2018} \\
        10 & IN3PD \cite{Faggioli:in3pd} & 0.2078 & 0.3713 & 1.9517 & \url{https://github.com/guglielmof/recsys_spt2018MI} \\\hline\hline
    \end{tabular}
    }
    \label{tab:main}
\end{table*}

\begin{table*}[t]
    \centering
    \caption{Final results achieved by the top 10 teams in the creative track. The highest R-prec and NDCG as well as the lowest clicks are marked as bold.}
    \scalebox{0.77}{
    \begin{tabular}{llcccl}\hline\hline
        Rank & Team & R-prec & NDCG & clicks & code \\ \hline
        1 & vl6 \cite{Volkovs:vl6} & \textbf{0.2234} & \textbf{0.3939} & \textbf{1.7845} & \url{https://github.com/layer6ai-labs/vl6_recsys2018}\\
        2 & Creamy Fireflies \cite{Antenucci:creamy-fireflies} & 0.2197 & 0.3846 & 1.9252 & \url{https://github.com/tmscarla/spotify-recsys-challenge} \\
        3 & KAENEN \cite{Ludewig:kaenen} & 0.2090 & 0.3746 & 2.0482 & \url{https://github.com/rn5l/rsc18} \\
        4 & cocoplaya \cite{Ferraro:cocoplaya} & 0.2022 & 0.3656 & 1.8377 & \url{https://github.com/andrebola/creative-recsys-cocoplaya} \\
        5 & BachPropagate \cite{Kallumadi:bach-propagate} & 0.2024 & 0.3659 & 2.0029 & \url{https://bachpropagate.weebly.com/} \\
        6 & Trailmix \cite{Zhao:trailmix} & 0.2059 & 0.3703 & 2.2589 & \url{https://github.com/xing-zhao/RecSys-Challenge-2018-Trailmix.git} \\
        7 & teamrozik \cite{Kaya:teamrozik} & 0.2054 & 0.3609 & 2.1636 & \url{https://github.com/mesutkaya/SpotifyRecSysChallenge2018} \\
        8 & Freshwater Sea & 0.1952 & 0.3504 & 2.1302 & \url{https://github.com/fyrelab/Spotify-RecSys} \\
        9 & Team Radboud \cite{Niedek:radboud} & 0.1982 & 0.3564 & 2.2934 & \url{https://github.com/TimovNiedek/recsys-random-walk} \\
        10 & spotif.ai \cite{Kim:spotifai} & 0.1924 & 0.3394 & 2.2665 & \url{https://github.com/eldrin/recsys18-spotify-spotif-ai} \\
        10 & Avito \cite{Rubtsov:avito} & 0.1764 & 0.3337 & 1.8988 & \url{https://github.com/VasiliyRubtsov/recsys2018} \\\hline\hline
    \end{tabular}
    }
    \label{tab:creative}
\end{table*}

\begin{table*}[t]
    \centering
    \caption{The performance of top 10 teams in the main track for different types of playlists in the challenge set. The highest R-prec and NDCG as well as the lowest clicks are marked as bold.}
    \renewcommand{\tabcolsep}{3pt}
    \scalebox{0.68}{ % 0.94
    \begin{tabular}{lccccccccccccccc} \hline\hline
        \multirow{3}{*}{Team} & \multicolumn{3}{c}{Type 1} & \multicolumn{3}{c}{Type 2} & \multicolumn{3}{c}{Type 3} & \multicolumn{3}{c}{Type 4} & \multicolumn{3}{c}{Type 5} \\ \cline{2-16}
        & \multicolumn{3}{c}{title only} & \multicolumn{3}{c}{title + first 5 tracks} & \multicolumn{3}{c}{only first 5 tracks} & \multicolumn{3}{c}{title + first 10 tracks} & \multicolumn{3}{c}{only first 10 tracks} \\ \cline{2-16}
        & R-prec & NDCG & clicks & R-prec & NDCG & clicks & R-prec & NDCG & clicks & R-prec & NDCG & clicks  & R-prec & NDCG & clicks \\ \hline
        vl6 & \textbf{0.0978} & 0.2044 & 10.759 & 0.2032	& 0.3766 & 0.900 & 0.2089 & 0.3847 & 0.644 & 0.2098 & \textbf{0.3973} & 0.437 & 0.1955 & 0.3737 & 0.631 \\
        hello world! & 0.0870 & 0.1925 & 11.400 & \textbf{0.2035} & \textbf{0.3791} & 0.908 & \textbf{0.2153} & \textbf{0.3939} & 0.646 & 0.2090 & 0.3928 & 0.465 & \textbf{0.1994} & \textbf{0.3788} & 0.571 \\
        Avito & 0.0845 & 0.1881 & \textbf{10.423} & 0.2008 & 0.375 & \textbf{0.875} & 0.2103 & 0.3878 & \textbf{0.623} & \textbf{0.2104} & 0.3956 & \textbf{0.424} & 0.1970 & 0.3752 & \textbf{0.568} \\
        Creamy Fireflies & 0.0949 & 0.1959 & 10.959 & 0.1979 & 0.3682 & 1.026 & 0.2123 & 0.3868 & 0.766 & 0.2034 & 0.3841 & 0.708 & 0.1968 & 0.3695 & 0.773\\
        MIPT\_MSU & 0.0948 & 0.1994 & 10.797 & 0.1946 & 0.3648 & 1.013 & 0.2061 & 0.3821 & 0.695 & 0.1940 & 0.3793 & 0.635 & 0.1895 & 0.3624 & 0.700\\
        HAIR & 0.0829 & 0.1812 & 12.932 & 0.1956 & 0.3660 & 1.000 & 0.2037 & 0.3740 & 0.756 & 0.2002 & 0.3810 & 0.534 & 0.1929 & 0.3655 & 0.640\\
        KAENEN & 0.0953 & \textbf{0.2053} & 10.563 & 0.1945 & 0.3611 & 1.168 & 0.2049 & 0.3776 & 1.039 & 0.1969 & 0.3754 & 0.759 & 0.1897 & 0.3615 & 0.961\\
        BachPropagate & 0.0751 & 0.1814 & 10.426 & 0.1991 & 0.3694 & 1.038 & 0.2070 & 0.3813 & 0.783 & 0.2034 & 0.3842 & 0.597 & 0.1940 & 0.3661 & 0.749\\
        Definitive Turtles & 0.0960 & 0.2001 & 10.884 & 0.1935 & 0.3651 & 1.212 & 0.2049 & 0.3797 & 0.893 & 0.1951 & 0.3755 & 0.769 & 0.1887 & 0.3623 & 0.946\\
        IN3PD & 0.0963 & 0.2031 & 10.452 & 0.1935 & 0.3608 & 1.108 & 0.2076 & 0.3813 & 0.753 & 0.1981 & 0.3772 & 0.573 & 0.1899 & 0.3615 & 0.746\\\hline\hline
        \multirow{3}{*}{Team} & \multicolumn{3}{c}{Type 6} & \multicolumn{3}{c}{Type 7} & \multicolumn{3}{c}{Type 8} & \multicolumn{3}{c}{Type 9} & \multicolumn{3}{c}{Type 10} \\ \cline{2-16}
        & \multicolumn{3}{c}{title + first 25 tracks} & \multicolumn{3}{c}{title + 25 random tracks} & \multicolumn{3}{c}{title + first 100 tracks} & \multicolumn{3}{c}{title + 100 random tracks} & \multicolumn{3}{c}{title + first track} \\ \cline{2-16}
        & R-prec & NDCG & clicks & R-prec & NDCG & clicks & R-prec & NDCG & clicks & R-prec & NDCG & clicks  & R-prec & NDCG & clicks \\ \hline
        vl6 & 0.2488 & 0.4005 & 0.262 & \textbf{0.3718} & \textbf{0.5616} & 0.024 & 0.1888 & 0.3539 & 0.634 & 0.3656 & \textbf{0.5846} & 0.019 & \textbf{0.1510} & \textbf{0.3087} & \textbf{3.529} \\
        hello world! & \textbf{0.2584} & \textbf{0.4138} & \textbf{0.144} & 0.3559 & 0.5417 & 0.021 & \textbf{0.2225} & \textbf{0.3956} & \textbf{0.319} & 0.3414 & 0.5538 & 0.033 & 0.1408 & 0.2901 & 4.445\\
        Avito & 0.2544 & 0.4082 & 0.162 & 0.3440 & 0.5340 & 0.022 & 0.2036 & 0.3728 & 0.438 & 0.3054 & 0.5162 & 0.081 & 0.1429 & 0.2927 & 4.202\\
        Creamy Fireflies & 0.2454 & 0.3921 & 0.260 & 0.3563 & 0.5384 & 0.037 & 0.2073 & 0.3691 & 0.507 & 0.3476 & 0.5611 & 0.035 & 0.1402 & 0.2916 & 4.264\\
        MIPT\_MSU & 0.2251 & 0.3755 & 0.404 & 0.3717 & 0.5540 & \textbf{0.017} & 0.1689 & 0.3276 & 0.888 & \textbf{0.3739} & 0.5754 & \textbf{0.014} & 0.1489 & 0.3029 & 3.591\\
        HAIR & 0.2388 & 0.3847 & 0.257 & 0.3558 & 0.5363 & 0.047 & 0.1952 & 0.3528 & 0.615 & 0.3611 & 0.5741 & 0.039 & 0.1366 & 0.2870 & 4.995\\
        KAENEN & 0.2370 & 0.3817 & 0.426 & 0.3375 & 0.5240 & 0.060 & 0.1886 & 0.3466 & 1.011 & 0.3060 & 0.5235 & 0.049 & 0.1402 & 0.2906 & 4.504\\
        BachPropagate & 0.2405 & 0.3872 & 0.293 & 0.3364 & 0.5171 & 0.038 & 0.1919 & 0.3501 & 0.778 & 0.3005 & 0.5084 & 0.035 & 0.1418 & 0.2944 & 4.097\\
        Definitive Turtles & 0.2366 & 0.3830 & 0.424 & 0.3342 & 0.5195 & 0.077 & 0.1931 & 0.3532 & 0.877 & 0.3062 & 0.5208 & 0.056 & 0.1377 & 0.2917 & 4.643\\
        IN3PD & 0.2426 & 0.3882 & 0.296 & 0.3080 & 0.4813 & 0.069 & 0.1911 & 0.3504 & 0.597 & 0.3163 & 0.5241 & 0.046 & 0.1341 & 0.2850 & 4.877\\\hline\hline
    \end{tabular}
    }
    \label{tab:main:detailed}
\end{table*}

\begin{table*}[t]
    \centering
    \caption{The performance of top 10 teams in the creative track for different types of playlists in the challenge set. The highest R-prec and NDCG as well as the lowest clicks are marked as bold.}
    \renewcommand{\tabcolsep}{3pt}
    \scalebox{0.68}{ % .94
    \begin{tabular}{lccccccccccccccc} \hline\hline
        \multirow{3}{*}{Team} & \multicolumn{3}{c}{Type 1} & \multicolumn{3}{c}{Type 2} & \multicolumn{3}{c}{Type 3} & \multicolumn{3}{c}{Type 4} & \multicolumn{3}{c}{Type 5} \\ \cline{2-16}
        & \multicolumn{3}{c}{title only} & \multicolumn{3}{c}{title + first 5 tracks} & \multicolumn{3}{c}{only first 5 tracks} & \multicolumn{3}{c}{title + first 10 tracks} & \multicolumn{3}{c}{only first 10 tracks} \\ \cline{2-16}
        & R-prec & NDCG & clicks & R-prec & NDCG & clicks & R-prec & NDCG & clicks & R-prec & NDCG & clicks  & R-prec & NDCG & clicks \\ \hline
        vl6 & \textbf{0.0979} & 0.2044 & 10.746 & \textbf{0.2032} & \textbf{0.3773} & \textbf{0.889} & 0.2084 & \textbf{0.3840} & \textbf{0.652} & \textbf{0.2094} & \textbf{0.3978} & \textbf{0.439} & \textbf{0.1949} & \textbf{0.3733} & 0.647\\
        Creamy Fireflies & 0.0946 & 0.1961 & 10.899 & 0.1978 & 0.3682 & 1.033 & \textbf{0.2095} & \textbf{0.3840} & 0.742 & 0.2019 & 0.3800 & 0.661 & 0.1919 & 0.3650 & 0.900 \\
        KAENEN & 0.0953 & \textbf{0.2053} & 10.563 & 0.1943 & 0.3617 & 1.172 & 0.2056 & 0.3776 & 1.046 & 0.1968 & 0.3754 & 0.752 & 0.1899 & 0.3616 & 0.958\\
        cocoplaya & 0.0724 & 0.1786 & \textbf{10.060} & 0.1877 & 0.3559 & 1.047 & 0.1962 & 0.3629 & 0.815 & 0.1954 & 0.3763 & 0.532 & 0.1824 & 0.3526 & 0.656\\
        BachPropagate & 0.0720 & 0.1794 & 10.662 & 0.1929 & 0.3607 & 1.173 & 0.2033 & 0.3761 & 0.956 & 0.1942 & 0.3747 & 0.672 & 0.1886 & 0.3599 & 0.943\\
        Trailmix & 0.0815 & 0.1817 & 12.638 & 0.1894 & 0.3585 & 1.124 & 0.2058 & 0.3798 & 0.889 & 0.1965 & 0.3776 & 0.749 & 0.1875 & 0.3608 & 0.957\\
        teamrozik & 0.0955 & 0.1959 & 11.363 & 0.1827 & 0.3405 & 1.522 & 0.1986 & 0.3592 & 0.868 & 0.1923 & 0.3604 & 0.730 & 0.1843 & 0.3482 & 0.873\\
        Freshwater Sea & 0.0885 & 0.1870 & 11.367 & 0.1837 & 0.3448 & 1.271 & 0.1985 & 0.3659 & 0.924 & 0.1800 & 0.3481 & 0.719 & 0.1761 & 0.3364 & 1.012\\
        Team Radboud & 0.0883 & 0.1951 & 12.853 & 0.1858 & 0.3455 & 1.340 & 0.1982 & 0.3658 & 0.903 & 0.1899 & 0.3627 & 0.683 & 0.1818 & 0.3469 & 0.786\\
        spotif.ai & 0.0720 & 0.1750 & 10.157 & 0.1674 & 0.3101 & 1.740 & 0.1778 & 0.3254 & 0.982 & 0.1742 & 0.3328 & 0.935 & 0.1679 & 0.3197 & 0.958\\
        Avito & 0.0800 & 0.1831 & 9.934 & 0.1634 & 0.3289 & 1.124 & 0.1672 & 0.3328 & 0.842 & 0.1772 & 0.3529 & 0.530 & 0.1616 & 0.3276 & \textbf{0.614}\\\hline\hline
        \multirow{3}{*}{Team} & \multicolumn{3}{c}{Type 6} & \multicolumn{3}{c}{Type 7} & \multicolumn{3}{c}{Type 8} & \multicolumn{3}{c}{Type 9} & \multicolumn{3}{c}{Type 10} \\ \cline{2-16}
        & \multicolumn{3}{c}{title + first 25 tracks} & \multicolumn{3}{c}{title + 25 random tracks} & \multicolumn{3}{c}{title + first 100 tracks} & \multicolumn{3}{c}{title + 100 random tracks} & \multicolumn{3}{c}{title + first track} \\ \cline{2-16}
        & R-prec & NDCG & clicks & R-prec & NDCG & clicks & R-prec & NDCG & clicks & R-prec & NDCG & clicks  & R-prec & NDCG & clicks \\ \hline
        vl6 & \textbf{0.2485} & \textbf{0.4006} & 0.265 & \textbf{0.3710} & \textbf{0.5603} & \textbf{0.023} & 0.1869 & 0.3523 & 0.636 & \textbf{0.3638} & \textbf{0.5825} & \textbf{0.021} & \textbf{0.1497} & \textbf{0.3065} & \textbf{3.527}\\
        Creamy Fireflies & 0.2454 & 0.3921 & 0.261 & 0.3534 & 0.5341 & 0.028 & \textbf{0.2081} & \textbf{0.3688} & \textbf{0.476} & 0.3517 & 0.5645 & 0.034 & 0.1427 & 0.2926 & 4.218\\
        KAENEN & 0.2373 & 0.3815 & 0.417 & 0.3372 & 0.5240 & 0.058 & 0.1886 & 0.3465 & 0.993 & 0.3060 & 0.5229 & 0.048 & 0.1392 & 0.2896 & 4.475\\
        cocoplaya & 0.2418 & 0.3886 & \textbf{0.154} & 0.3254 & 0.5090 & 0.034 & 0.1884 & 0.3439 & 0.619 & 0.2992 & 0.5069 & 0.065 & 0.1331 & 0.2812 & 4.395\\
        BachPropagate & 0.2354 & 0.3812 & 0.368 & 0.3189 & 0.5001 & 0.040 & 0.1885 & 0.3435 & 1.085 & 0.2897 & 0.4935 & 0.055 & 0.1402 & 0.2897 & 4.075\\
        Trailmix & 0.2407 & 0.3905 & 0.309 & 0.3352 & 0.5151 & 0.046 & 0.1863 & 0.3492 & 0.854 & 0.3049 & 0.5086 & 0.034 & 0.1313 & 0.2815 & 4.989\\
        teamrozik & 0.2380 & 0.3774 & 0.210 & 0.3340 & 0.5024 & 0.024 & 0.1862 & 0.3394 & 0.715 & 0.3113 & 0.5047 & 0.041 & 0.1316 & 0.2808 & 5.290\\
        Freshwater Sea & 0.2268 & 0.3635 & 0.196 & 0.2984 & 0.4663 & 0.061 & 0.1859 & 0.3373 & 0.730 & 0.2845 & 0.4753 & 0.085 & 0.1296 & 0.2796 & 4.937\\
        Team Radboud & 0.2267 & 0.3652 & 0.337 & 0.3130 & 0.4877 & 0.080 & 0.1759 & 0.3225 & 0.874 & 0.2865 & 0.4881 & 0.069 & 0.1358 & 0.2841 & 5.009\\
        spotif.ai & 0.2098 & 0.3363 & 0.427 & 0.3397 & 0.5156 & 0.024 & 0.1627 & 0.3044 & 0.919 & 0.3344 & 0.5416 & 0.024 & 0.1186 & 0.2334 & 6.499\\
        Avito & 0.2171 & 0.3580 & 0.239 & 0.2786 & 0.4432 & 0.110 & 0.1735 & 0.3287 & 0.617 & 0.2366 & 0.4296 & 0.195 & 0.1083 & 0.2524 & 4.783\\\hline\hline
    \end{tabular}
    }
    \label{tab:creative:detailed}
\end{table*}

\section{Top-performing approaches: Main Track}
\label{sec:winners_main}
In this section, we provide a brief analysis of the approaches taken by the top 10 teams in the main track. We further explain the approaches used by the top 3 teams in more detail. 

High-level characteristics of the winning approaches are presented in \tablename~\ref{tab:winning=methods}. As shown in the table, several teams took advantage of a two-stage architecture for the playlist continuation task. In such an architecture, the first stage model retrieves a small set of tracks (compared to the total number of tracks in the dataset), while the second stage focuses on re-scoring or re-ranking the output of the first stage model with the goal of accuracy improvement. Therefore, a high-recall model is desired for the first stage, however, a high-precision model is preferred for the second stage. The reason for making this decision is mainly related to efficiency. However, the two-stage architecture can also improve the APC performance. Among the top 10 teams in the main track, vl6 \cite{Volkovs:vl6}, Avito \cite{Rubtsov:avito}, HAIR \cite{Zhu:hair}, BachPropagate \cite{Kallumadi:bach-propagate}, and IN3PD \cite{Faggioli:in3pd} took advantage of a multi-stage architecture. Multi-stage models have been extensively explored for improving efficiency and effectiveness in various retrieval and recommendation settings \cite{Chen:2017,Dang:2013,Lampropoulos:2012,Li:2011,Wang:2011}.

In addition, matrix factorization, as a dominant approach in collaborative filtering (CF), was also employed by several top performing teams, including vl6 \cite{Volkovs:vl6}, Avito \cite{Volkovs:vl6}, KAENEN \cite{Ludewig:kaenen}, and IN3PD \cite{Faggioli:in3pd}. These models mostly create an incomplete playlist-track matrix and use matrix factorization to learn a low-dimensional dense representation for each playlist and track. They learn similar representations for the tracks that often occur together in user-created playlists. Therefore, the tracks from a single artist (band), an album, or a music genre may be assigned close representations. The matrix factorization algorithms used by the top teams include weighted regularized matrix factorization (WRMF) \cite{Hu:2008}, LightFM with a weighted approximate-rank pairwise (WARP) loss \cite{Kula:2015}, and Bayesian personalized ranking (BPR) \cite{Rendle:2009}. Interestingly, some teams, including HAIR \cite{Zhu:hair} and Definitive Turtles \cite{Kelen:definitive-turtles}, were able to achieve promising results using simple neighborhood-based collaborative filtering methods.

Moreover, due to the high capacity of neural networks to learn task-specific representations, a number of top performing teams used neural network models to produce accurate predictions for the APC task. These neural approaches include: (1) simple feed-forward networks for predicting tracks given each playlist (e.g., a word2vec-style model \cite{Mikolov:2013}) or for neural collaborative filtering \cite{He:2017}, (2) convolutional models for playlist embedding or extracting useful information from playlist titles, (3) recurrent neural networks and in particular long short-term memory networks for modeling the sequence of tracks in the playlists, and (4) autoencoders for learning playlist representations. 

Most top-performing teams that used a two-stage architecture built their second stage based on (mostly pairwise) learning to rank models. These models were designed to re-rank a small number of tracks given a set of features produced by different models, including the first-stage model, as well as several heuristic hand-crafted features. The tree-based models, such as XGBoost \cite{Chen:2016}, GBDT \cite{Friedman:2001}, and LambdaMART \cite{Burges:2010}, were the popular learning to rank algorithms among the top teams in the challenge.

It is notable that some top performing teams used information retrieval techniques mainly developed for the ad-hoc retrieval task. For instance, inverse document frequency (IDF) weighting \cite{Jones:1972}, TF-IDF weighting \cite{Salton:1988}, BM25 weighting \cite{Robertson:1994}, and relevance model \cite{Lavrenko:2001} (a pseudo-relevance feedback model) were respectively employed by teams Definitive Turtles \cite{Kelen:definitive-turtles}, KAENEN \cite{Ludewig:kaenen}, Creamy Fireflies \cite{Antenucci:creamy-fireflies}, and BachPropagate \cite{Kallumadi:bach-propagate}.

An important challenge in the APC task is dealing with cold-start playlists, i.e., the playlists with only title (no track). Some teams tried to deal with such special cases differently by trying to learn a relationship between the playlist titles and its tracks. Among which, neural networks and matrix factorization models are notable that predict the tracks in a playlist, given its title. 

In the following, we detail the approaches taken by the top three teams in the main track:

\textbf{vl6 team:} The vl6 team used a two-stage architecture, where the first one is based on Weighted Regularized Matrix Factorization (WRMF) \cite{Hu:2008}, and the second one is implemented using XGBoost \cite{Chen:2016}, a gradient boosting learning to rank model. In addition to the output of the WRMF model, few models were used to produce features for the XGBoost model. These models include a convolutional neural network for playlist embedding, user-user and item-item neighborhood-based collaborative filtering models, and a set of hand-crafted features. Note that the cold-start instances (those that only consists of a title with no track) were handled separately. For such cases, the vl6 team used a matrix factorization on top of the playlist titles. For a detailed description of the approach used by the vl6 team, refer to \cite{Volkovs:vl6}. 

\textbf{hello world! team:} The team hello world! linearly combined the results produced by two different models: an autoencoder model and a convolutional neural network. The autoencoder model tries to reconstruct track lists and artist lists for each playlist. To model both marginal and joint information across playlist and contents, the model was trained using a ``hide-and-seek'' idea. In other words, either the track list or the artist list was randomly deactivated in the input of the autoencoder. To use the title of playlist, especially for the cold-start situations, a character-level convolutional neural network (charCNN) was used to learn a representation from the playlist's title. This can be viewed as a classification model: predicting the tracks in each playlist given its title. In the linear combination, the output of the charCNN was weighted higher for shorter playlists. For a detailed description of the approach used by the team hello world!, we refer the reader to \cite{Yang:hello-world}. 

\textbf{Avito team:} Similar to the first team, the team Avito also used a two-stage architecture. The first stage is based on a matrix factorization model with the weighted approximate-rank pairwise (WARP) loss, implemented in LightFM \cite{Kula:2015}. Two separate models were trained, one based on playlist-track information and the other one based on the playlist titles. The union of the outputs of these two models were re-ranked by the second stage model, which is a XGBoost learning to rank model \cite{Chen:2016}. In addition to the LightFM features, some additional feature engineering was done to boost the performance. For a detailed description of the approach used by the Avito team, refer to \cite{Rubtsov:avito}.

\begin{table}[t]
    \centering
    \caption{Characteristics of top-performing approaches in the main track. Two stage, MF, NN, and LTR denote two-stage cascaded architecture, matrix factorization, neural networks, and learning to rank, respectively.}
    \begin{tabular}{llcccc} \hline\hline
        rank & team & two stage & MF & NN & LTR  \\\hline
        1 & vl6 & \cmark & \cmark & \cmark & \cmark  \\
        2 & hello world! & \xmark & \xmark & \cmark & \xmark  \\
        3 & Avito & \cmark & \cmark & \xmark & \cmark  \\
        4 & Creamy Fireflies & \xmark & \xmark & \xmark & \xmark  \\
        6 & HAIR & \cmark & \xmark & \cmark & \cmark  \\
        7 & KAENEN & \xmark & \cmark & \xmark & \xmark  \\
        7 & BachPropagate & \cmark & \xmark & \cmark & \cmark  \\
        9 & Definitive Turtles & \xmark & \xmark & \xmark & \xmark \\
        10 & IN3PD & \cmark & \cmark & \xmark & \xmark \\\hline\hline
    \end{tabular}
    \label{tab:winning=methods}
\end{table}

\section{Top-performing Approaches: Creative Track}\label{sec:winners_creative} % External Resources for Playlist Continuation: 
In this section, we provide a brief analysis of the approaches taken by the top 10 teams in the creative track, in which teams were allowed to use external resources.\footnote{When teams started to submit the same approaches to the creative and main tracks (due to the lower popularity of the creative one), we \emph{required} submissions to the creative track to exploit external data.} We further explain the approaches followed by the top 3 teams in more detail. 

A first observation when reviewing the algorithms of the top performers in the creative track reveals that most of the teams only slightly altered their algorithms for the main track, e.g., by adding to their pipeline a final audio content-based re-ranking approach \cite{Ludewig:kaenen} or by extending their content-based filtering approaches by enriching the provided meta-data with audio information \cite{Antenucci:creamy-fireflies}.
Most of what was said above for the main track therefore also holds for the approaches taken in the creative track, in particular the superior performance of two-stage architectures, use of neural networks, and special handling of cold-start situations.

Interestingly, except for one team (spotif.ai), all top 10 teams participating in the creative track also participated in the main track (see Table~\ref{tab:creative}). However, their ranks most often differed between the main and creative tracks: vl6 (ranked 1st in main track), Creamy Fireflies (4th in main), KAENEN (7th in main), cocoplaya (11th in main), BachPropagate (7th in main), Trailmix (13th in main), teamrozik (63rd in main), Freshwater Sea (19th in main), Team Radboud (21st in main), and Avito (3rd in main).
The spotif.ai team, which solely participated in the creative track, employed a recurrent neural network architecture (long short-term memory~\cite{doi:10.1162/neco.1997.9.8.1735}) that was particularly designed to cope with sequential data, in addition to a weighted regularized matrix factorization (WRMF) approach~\cite{Hu:2008}.

Remarkably, almost all teams participating in the creative track used the Spotify API\footnote{\url{https://developer.spotify.com/documentation/web-api/reference/tracks/get-audio-features}}
 as external data source and downloaded the provided audio content features.
A notable exception was team cocoplaya \cite{Ferraro:cocoplaya}, who retrieved 30-second-snippets of each track from Spotify and computed their own audio-based features, in particular the output of a probabilistic genre classifier for each of 13 genres~\cite{bogdanov_etal:ismir:2016}. Others included external information when filtering playlist titles using stopword lists or pre-defined lists of music-related terms (e.g., playlist, songs, music) \cite{Zhao:trailmix}.
Still others used pre-trained word embedding models, such as the CBOW model from word2vec~\cite{Mikolov:2013}, to create track embeddings \cite{Kallumadi:bach-propagate}.

% description of innovative (not necessarily winning) approaches
%\ms{t.b.c.: description of general approaches and directions chosen by top 10 teams.}
%Especially in cold-start situations, the use of external data is vital.

In the following, we detail the approaches taken by the top three teams in the creative track:

\textbf{vl6 team:} The vl6 team also ranked first in the creative track. Their approach taken here largely resembles the one taken in the main track (see Section~\ref{sec:winners_main}). The only difference is that the feature set used in the second stage of their approach (feature selection using an XGBoost model) was extended by content-based music descriptors of tracks. These descriptors were acquired through the Spotify Audio API and comprise acousticness, danceability, energy, instrumentalness, key, liveness, loudness, mode, speechiness, tempo, time signature, and valence. However, no substantial and consistent improvement was achieved by adding these features (compare Tables~\ref{tab:main} and~\ref{tab:creative}).
For a detailed description of the approach used by the vl6 team, refer to \cite{Volkovs:vl6}. 

\textbf{Creamy Fireflies team:} 
This team used an ensemble of known techniques, which they intelligently combined in an informed way to select and tune the individual techniques depending on the underlying playlist characteristics (from only title to 100 tracks).
%a series of boosts to be applied on top of the final predictions and improve the recommendation quality.
Five base approaches were used: (1) popularity-based recommendation, (2) track- and (3) playlist-based collaborative filtering (on the playlist-track matrix), as well as (4) track- and (5) playlist-based content-based filtering; (4) using artist and album identifiers as features; (5) additional features derived from playlist titles.
More precisely, playlist features were created by applying techniques from information retrieval and natural language processing to clean and enrich the playlist titles (e.g., tokenization, normalization, and stemming).
%to address cold start in the scenario where only playlist titles were given, this team applied techniques
In a tuning step, the authors then sought optimal parameters for each combination of algorithm and playlist category (cf.~Section~\ref{sec:data}). 
%\ms{@Hamed, we should in the beginning of the paper, clearly report all 10 categories of playlists (not only in the result tables). I think this is crucial for understanding. And also we can refer the reader to the section where the 10 categories are explained.} \hz{see Section \ref{sec:data}, is it satisfying?}) % PERFECT!
Their base ensemble approach subsequently %The team then designed different ensemble approaches for playlists exhibiting different characteristics. In a base ensemble, they optimized and 
weighted the five algorithms for each playlist category and other playlist characteristics (e.g., length and track positions). The final score was computed as the weighted sum of the scores given by each algorithm and playlist category.
The authors also investigated another ensemble model, based on a proposed measure of artist heterogeneity. Clustering 
the playlists according to this measure and performing a cluster-based filtering slightly improved NDCG and R-precision.
Eventually, several boosts depending on the playlist category were investigated. For instance, assuming that the last tracks in a (long) seed playlist are the most important ones with respect to the continuation, candidate tracks more similar to those last ones in the seed playlist were given higher weight.

In the creative track, team Creamy Fireflies additionally used the Spotify API to acquire the following features for each track: acousticness, danceability, energy, instrumentalness, liveness, loudness, speechiness, tempo, valence, and popularity.
They extended their content-based filtering and collaborative filtering models described above to include track-level similarity. To this end, a sparse representation of track clusters was used, in which clusters were generated by grouping tracks into four equally sized clusters based on the values of each audio feature.
For a detailed description of the approach used by the Creamy Fireflies team, refer to \cite{Antenucci:creamy-fireflies}. 

\textbf{KAENEN team:} 
Also the KAENEN team proved that it is possible to achieve remarkable results without using very complex approaches. They combined nearest-neighborhood techniques with common matrix factorization algorithms, which were adapted to the application domain. %and finally applied several heuristics.
More precisely, they adapted an item-based CF approach, treating playlists as users and computing cosine similarity between item vectors (binary, over all playlists). To alleviate the popularity bias that affects such co-occurrence-based similarities, inverse document frequency (IDF) weighting is applied to each candidate track, i.e., tracks that appear in many playlists are downweighted.
As second approach, the team proposed a playlist-based nearest neighbor method, which uses the same framework as the item-based CF approach, but this time computing similarities over binary playlist vectors instead of track vectors. Each candidate track $t$ is then ranked with respect to the similarity to the most similar playlists in which $t$ occur, again considering the IDF weighting.
As third approach, the team adapted a standard matrix factorization technique using alternating least squares (ALS) optimization. To compute the ranking of a candidate track $t$ with respect to a seed playlist $p$, the latent factors of all tracks in $p$ are IDF-weighted and the dot product of the arithmetic mean of this set of latent factors (constituted of all tracks in $p$) and the latent factors of $t$ is used as final score.
To address the cold-start scenario (only playlist title given), the team used a simple string matching technique applied on tokenized and stemmed playlist titles to identify the most similar playlists to $p$. In addition, they used a matrix factorization approach (with ALS optimization) treating unique playlist names as users and occurrences of tracks in the corresponding playlists as ``ratings''. The latent factors were then used to identify the playlists most similar to $p$.
The individual approaches described above were subsequently combined into a hybrid recommender system, using switching and weighting hybridization schemes~\cite{Burke2002}. In cold-start cases where the string matching approaches did not produce enough results (i.e., 500 tracks), the missing ones were filled with the most popular tracks of the MPD.

For the creative track, like the other top performers, the team KAENEN retrieved audio features using the Spotify API.
They then used a re-ranking strategy as follows. If the mean standard deviation of the audio features of the seed playlist $p$'s tracks fell below a threshold (low content diversity), the original score of a candidate track $t$ with respect to $p$ was re-weighted by cosine similarity between $t$'s content features and the mean of the content features of all tracks in $p$.
For a detailed description of the approach used by the KAENEN team, refer to~\cite{Ludewig:kaenen}. 

\section{Other Notable Approaches}\label{sec:other_approaches}
In the previous sections, we discussed the approaches of the top teams in each of the challenge tracks. A detailed analysis of all 117 active teams' approaches is unfeasible, due to the sheer number of teams, as well as the fact that only some of them published their approach in detail, or had sufficient documentation in the code they shared (with many teams not sharing their code at all). However, based on a review of some of the teams that did not achieve top scores, we see a similar variety of techniques used as in the top performing submissions. Some combination of collaborative filtering, word embedding approaches, deep neural network architectures, information retrieval techniques and ensembles thereof are used by teams who achieved both higher and lower scores. This raises the question of what makes one approach score better at the task than another? We can expect implementation details such as hyperparameter tuning, dataset preprocessing and sampling strategies to have a significant impact on the performance of an approach. Different formulations of objective functions, different approaches to extracting features from the dataset, as well as different architectures and sequencing of operations could also have an effect on the overall results.
To provide some context towards answering this question, we present two teams which did not achieve scores in the top 10, but which took different approaches to solving the automatic playlist continuation task:

\textbf{Unconscious Bias team:}
The Unconscious Bias placed 43rd in the main track. Their approach is based on applying adversarial autoencoders~\cite{makhzani2015adversarial} to the playlist continuation task. On the surface, this approach shares similarities to the approach taken by team hello world!, which came in 2nd place in the main track. Team hello world!~\cite{Yang:hello-world} used a combination of a content-aware autoencoder as well as a convolutional neural network on playlist titles to arrive at their score. 
In contrast to hello world!'s various novel dropout strategies to train an autoencoder network, the Unconscious Bias team uses an adversarial approach as a regularization technique, which allows the network to generalize from the training set to unseen examples, in a way that also matches the prior distribution. Interestingly, Unconscious Bias evaluated the general autoencoder approach as a baseline in their experiments, and found performance to be lower than their proposed adversarial autoencoder approach. 
Clearly there are very significant differences in the two approaches, even though both utilize autoencoders. To delve deeper into these differences and how they might have resulted in such a large difference in scores, we recommend reading both ~\cite{Vagliano:unconscious-bias} and ~\cite{Yang:hello-world}.

\textbf{D2KLab team:}
Like many other teams, D2KLab took an ensemble approach to the problem, combining several methods together to solve the task, including a specialized method to handle the cold-start (title-only) use case. Their core approach involves an ensemble of multiple Recurrent Neural Networks (RNN), in particular, Long-Short Term Memory (LSTM) cells trained to predict the next track given a sequence of tracks. The inputs to the system are word2vec embeddings at the track, album, and artist level. To deal with playlist titles, and particularly to address the cold-start use case, they also derived title embeddings using the fastText~\cite{bojanowski2016enriching} algorithm, trained on n-grams of playlist titles included in groups of playlists that are clustered in the playlist embedding space.

For their creative track submission, D2KLab also included lyric metadata by linking the MPD tracks with the WASABI lyric corpus~\cite{meseguer-brocal-wasabi}. They developed a suite of lyric features that describe the different stylistic and linguistic dimensions of a song text, for example, vocabulary and emotion.  These features were vectorized and concatenated with the other embedding-based features as inputs to the RNN network.

In the main track, their submission achieved an R-Precision of	0.1808, NDCG of 0.3252, and Clicks score of 3.086, which ranks them in the 37th position. In the creative track, their approach achieved an R-Precision of 0.1852, NDCG of 0.3334, and Clicks score of 3.026, putting them in 13th place. The improved scores in the creative track  suggests that their use of lyric features adds valuable information for the playlist continuation task. For a detailed description of the approach used by the D2KLab team, refer to ~\cite{Monti:d2klab}.

\section{Summary of Key Findings}
% \hz{who: Paul}
In this section, we briefly summarize our key findings from the challenge and the submitted solutions. In summary, most approaches ensemble the results obtained by several well-known methods, including matrix factorization models, neighborhood-based collaborative filtering models, basic information retrieval techniques, and learning to rank models. The results show that the models work best when a sufficient number of tracks per playlist is provided and they are randomly selected from the playlist (as opposed to the sequential order from the beginning of the playlist). The submitted solutions could not effectively use playlist titles for APC. This might be due to the sparseness of the titles as well as the scale of the training data. In addition, none of the submitted solutions tried to infer the user intents from the playlist titles. The results also demonstrate that the performance of different models are close to each other when few tracks per playlist are given. However, when the number of tracks increases, a more diverse set of results is observed.

In the creative track, most teams exclusively used the descriptors from the Spotify API, and only few of them tried to extract their own features from the audio. It is worth noting that surprisingly, there is no significant gap between the results in the main and creative tracks. Indeed, the results for the creative track are marginally worse than those obtained for the main track. This might be due to the fact the inclusion of side information makes the problem more complex and the submitted solutions could not successfully generalize the information obtained from the exploited external resources.

% \hz{elaborate more}
% summarizing our findings during the challenge. what are the useful approaches. what we learned? are they generalizable? 

% \ms{some ideas: offline evaluation could be complemented in the future by additional, alternative measures of perceived utility, coherence, serendipity, etc.}

% \ms{trying to infer the user intent from playlist titles has not been tried}

% \ms{surprisingly(?), in the creative track, teams almost exclusively used descriptors from the Spotify API; only few teams tried to download audio and computed their own audio features}

% \ms{performance results for main and creative tracks were not substantially different. surprisingly, those for the creative track were even marginally lower than those of the main track: inclusion of side information didn't work?}

\section{Generalizability of Approaches and Results}

The RecSys Challenge 2018 focused on the topic of sequence-aware music recommendation and was deliberately and necessarily a narrow and clearly defined task (playlist continuation) as usual for such a competition.
Nevertheless, some of the best-performing approaches are transferable to target domains other than music, though to different extent which also depends on the target domain. 
Most straightforward, the approaches submitted to the \textit{main track}, which therefore do not use any external side information, could be adapted easily to multimedia domains such as (short) video, where users of platform like Youtube create and share their playlists of \textit{video clips}. Likewise, in the online learning and training domains, curated sequences of \textit{exercises} or \textit{tasks} are made available by teachers and students. Both share similar characteristics in the sense that the sequence of items does matter and consumption times are comparable in magnitude to those of songs.
Both factors, i.e., importance of sequences and similar consumption time~\cite{Schedl2018}, may prevent the immediate applicability of these approaches to other targets such as story lines of images (much shorter consumption time) or book reading lists (much longer consumption times and sequence often not important).

Nonetheless, for such domains which are further away from music, other ways of adopting the proposed approaches might be viable.
The models constructed from the provided dataset by some teams, most notably the two top performing ones in the main track (vl6 and hello world!) which are based on deep neural networks, could be used in a transfer learning setting to re-purpose the model for related tasks~\cite{Goodfellow2016}.

Most solutions submitted to the \textit{creative track} are harder to generalize, in particular if they are closely tied to content-based features. However, the level of generalizability obviously depends on the nature of the leveraged content features, which were used at the song and at the playlist level. 
Noteworthy, all top 3 teams (and many others) in the creative track used the Spotify API to extract audio descriptors (tempo, loudness, danceability, etc.).
As an example, team Creamy Fireflies relied in the creative track on artist and album identifiers, but also on Spotify's audio content descriptors to implement content-based filtering. While the former (identifiers) are practically available in almost all other domains too, audio content features are limited to a few domains (e.g., podcasts or videos).

As for the achieved results in terms of performance metrics, they strongly depend on the dataset used and vary according to the type of playlist in the challenge set on which they are computed. R-precision, NDCG, and number of clicks are therefore not comparable to results achieved on similar tasks in domains other than music. We are also not aware of existing research works or benchmarking challenges that easily compare to the RecSys Challenge 2018 in terms of the nature of the dataset and the distinction between different types of input playlists used in the evaluation of approaches. A detailed investigation of approaches and achievable results in other target domains using different kinds of playlists and target items therefore remains an avenue for future research.

Another avenue for generalization is given by the fact that the problem for playlist type 1 (title only) % and 10 (title and first track only) 
resembles a standard search or retrieval task, in which the query is expressed as text, i.e., the name of the playlist to create.
Successful approaches, taken in the RecSys Challenge 2018, which particularly address playlists of this type could therefore lead to improved capabilities to search and retrieve music by arbitrary natural language input. This would complement the current research on text-based music retrieval, which most often leverages (user-generated or expert-created) annotations or tags.
%It could be convincing to include a few more unique/interesting playlist titles and show the top results for those titles (I realize you may not have the raw data, or time, to do this, just an idea)
%There may also be a story around how this could generalize to Search (especially since we are using Search/IR-type metrics like precision and NDCG), especially if you consider the "Type 0" queries of title-only: research in this area could lead to improved capabilities to search and retrieve music by arbitrary terms, similar in a way to research done on tags (e.g. this work) It could be convincing to include a few more unique/interesting playlist titles and show the top results for those titles (I realize you may not have the raw data, or time, to do this, just an idea)

%The models that we've built at Spotify on top of this type of dataset have been invaluable as a general purpose similarity/recommendation technology: item-to-item, user-to-item. I'm not sure how we could show that with evidence (that sounds like actual research work) but perhaps if we can somehow explain that the task was necessarily narrow but that the use cases are broad, that might help? Maybe if we can identify other examples of "transfer learning" in this field, training on a specific task/dataset but then using it for general recommendation tasks, that might be useful?

\section{Future Directions and Open Avenues}
%\hz{who: all (streamline by Markus)}
Even though the RecSys Challenge 2018 has stimulated a wealth of ideas and creative solutions, we contemplate several directions for additional research that might be worth pursuing.

\paragraph{Integration of additional content and context feature:}
Given that solutions in the creative track did not outperform those in the main track, the question arises whether the right or good external data sources have been exploited by the algorithms submitted to the creative track. Almost all submissions relied on content features provided by the Spotify API, omitting the time-consuming task of computing other (maybe better) content descriptors from audio (snippets) of the tracks. Also additional contextual information about tracks, albums, or artists, e.g., Wikipedia articles or album reviews, could be integrated in the future.

\paragraph{Explicit inference of intent or purpose:}
In cases where a playlist title is given, sophisticated natural language processing techniques (NLP) could be applied, trying to uncover the listener's intent or purpose of the playlist. 
However, identifying such user intents to listen to music, the most important of which are arousal and mood regulation, achieving self-awareness, and expressing social relatedness~\cite{schaefer:frpsy:2013}, is challenging.
Therefore, NLP techniques will likely have to be complemented by insights gained from  gratification~\cite{lonsdale11bjp} and other psychological theories.

%first because of the broad range of intents~\cite{schaefer:frpsy:2013,lonsdale11bjp}, but also because of the diversity in the underlying playlist characteristics that may be considered to infer those intents~\cite{Schedl2018}. 
%Varying availability and amounts of information (e.g., playlist length and duration, availability of playlist title, or presence of contextual information) for different playlists or users likewise represent a challenge.

\paragraph{Modeling and transferring sequence-specific characteristics:}
We also see great potential for future approaches that analyze and model certain sequence-specific characteristics of user-generated playlists, formalize them, and integrate them into the sequential recommendation process. 
Similarly to the artist heterogeneity measure proposed by team Creamy Fireflies ~\cite{Antenucci:creamy-fireflies}, aspects of overall playlist coherence (e.g., in terms of genre, style, or acoustic descriptors), coherence of direct song-to-song transitions, or item diversity measures could be computed from user-generated playlists and considered as (weak) constraint in the process of APC, i.e., the seed playlist should be continued in a way that maintains the same level of coherence, diversity, etc.

\paragraph{Evaluation in terms of perceived recommendation quality:}
In addition to the mostly accuracy-related performance measures used to gauge performance of submissions, user-centric measures of perceived recommendation quality should be adopted in the future, in order to obtain a truly user-centric perspective of recommendation quality.
Such measures of perceived recommendation quality can be assessed through questionnaires in online evaluation settings. Existing questionnaires such as~\cite{Ekstrand2014UPD,Knijnenburg2012} should be extended to the sequence-aware music domain and may eventually include aspects of perceived accuracy, diversity, coherence, satisfaction, novelty, serendipity, and level of personalization.

%\ms{my view: (1) integration of other or more decent content features, contextual information, etc.; (2) explicitly inferring intent or purpose;  (3) this is related to understanding aspects of coherence, diversity, novelty, serendipity, etc. in user-generated playlists (and applying it to computer-generated ones); (4) evaluation (more than IR/ML/accuracy related ones))}

\section{Acknowledgements}
We would like to thank everyone at Spotify who was involved in the RecSys Challenge, including Ben Carterette, Christophe Charbuillet, Cedric de Boom, Jean Garcia-Gathright, James Kirk, James McInerney, Vidhya Murali, Hugh Rawlinson, Sravana Reddy, Marc Romejin, Romain Yon, and Yu Zhao. Furthermore, we greatly appreciate the help provided by previous organizers of the RecSys Challenge, in particular by Yashar Deldjoo, Mehdi Elahi, and Alan Said.

This work was supported in part by the Center for Intelligent Information Retrieval. Any opinions, findings and conclusions or recommendations expressed in this material are those of the authors and do not necessarily reflect those of the sponsor. %\hz{@Markus: add your funding, if any.} % None.

 {\scriptsize
 \bibliographystyle{abbrv}
%  \bibliography{RSC_2018}

 }

\appendix
%\counterwithin{figure}{section}
%\renewcommand{\thefigure}{\hbAppendixPrefix\arabic{figure}}
%\setcounter{figure}{5}

\section{Evaluation Metrics}
\label{appendix:eval}
As mentioned earlier in Section~\ref{sec:eval}, the quality of submissions were assessed based on the value of three different evaluation metrics: R-precision, normalized discounted cumulative gain (NDCG), and recommended songs clicks. In this appendix, we provide in detail description of each of these metrics.

\begin{itemize}[leftmargin=*]
    \item \textbf{R-precision} measures the fraction of recommended relevant items among all known relevant items (i.e., the number of withheld tracks) and is invariant of the order in which tracks are retrieved. R-precision is calculated on both the track and the artist level, with artist matches contributing a partial score (of 0.25) even if the predicted track is incorrect. Let $G_T$ and $G_A$ be the set of unique track IDs and artist IDs in the ground truth, respectively.  Let $S_T$ be the set of track IDs in the top $|G_T|$ tracks recommended in the submitted playlist, and $S_A$ be the set of unique artist IDs in the same set.  Then:
    \begin{align*}
        \text{R-precision} &= \frac{|S_T \cap G_T| + 0.25\cdot |S_A \cap G_A|}{|G_T|}
    \end{align*}
    The higher the R-precision, the better.
    
    \item \textbf{NDCG} \cite{Jarvelin:2002} assesses the ranking quality of the recommended tracks and increases when relevant tracks are placed higher in the recommendation list. This metric was originally proposed to evaluate the effectiveness of information retrieval systems. Nowadays, it is also frequently used for evaluating (music) recommender systems. Assuming that tracks for each playlist are sorted according to their recommendation score in descending order, the discounted cumulative gain (DCG) is then defined as follows:
    \begin{equation*}
        \text{DCG} = \sum_{i=1}^N \frac{r_{i}}{\log_{2} (i+1)}
    \end{equation*}
    where $r_{i}$ is the label (as found in the ground truth) for the item ranked at position $i$ for the playlist, and $N$ is the length of the recommendation list (here, $N=500$). DCG is normalized by IDCG -- the DCG value for the best possible ranking obtained by ordering the tracks by true ratings in descending order. NDCG is then calculated as:
    \begin{equation*}
        \text{NDCG} = \frac{\text{DCG}}{\text{IDCG}}
    \end{equation*}
    The higher the NDCG, the better.
    
    \item \textbf{Recommended songs clicks} (or shortly just ``clicks'') is a user-centric beyond-accuracy measure that relates to a Spotify feature called Recommended Songs. Given a playlist title and/or set of tracks in a playlist, this feature recommends 10 tracks to add to the playlist. The list can be refreshed to produce 10 more tracks. The recommended songs clicks metric is the number of refreshes needed before the first relevant track is encountered. It is formalized as shown in the following equation, where $R$ is the list of recommended tracks and $G$ is the ground truth, i.e., the omitted tracks from the real playlist.
    \begin{equation*}
        \text{clicks} =  \Big\lfloor \frac{\arg \min_i \{ R_i\!: R_i \in G \} - 1}{10} \Big\rfloor
    \end{equation*}
    If there is no relevant track in $R$, a value of 51 is picked, which is 1 plus the maximum number of clicks possible. The lower the recommended songs clicks, the better.
\end{itemize}

\section{Sample Playlist from the Dataset}
\label{appendix:sample_playlist}
A sample truncated playlist from the MDP dataset is presented below. %in Fig.~\ref{sampleplaylist}.

% \noindent\begin{minipage}{\textwidth}
\begin{lstlisting}[language=json,firstnumber=1,caption={A truncated sample playlist from MPD.},label=sampleplaylist]
{
    "name": "musical",
    "collaborative": "false",
    "pid": 5,
    "modified_at": 1493424000,
    "num_albums": 7,
    "num_tracks": 12,
    "num_followers": 1,
    "num_edits": 2,
    "duration_ms": 2657366,
    "num_artists": 6,
    "tracks": [
        {
            "pos": 0,
            "artist_name": "Degiheugi",
            "track_uri": "spotify:track:7vqa3sDmtEaVJ2gcvxtRID",
            "artist_uri": "spotify:artist:3V2paBXEoZIAhfZRJmo2jL",
            "track_name": "Finalement",
            "album_uri": "spotify:album:2KrRMJ9z7Xjoz1Az4O6UML",
            "duration_ms": 166264,
            "album_name": "Dancing Chords and Fireflies"
        },
        {
            "pos": 1,
            "artist_name": "Degiheugi",
            "track_uri": "spotify:track:23EOmJivOZ88WJPUbIPjh6",
            "artist_uri": "spotify:artist:3V2paBXEoZIAhfZRJmo2jL",
            "track_name": "Betty",
            "album_uri": "spotify:album:3lUSlvjUoHNA8IkNTqURqd",
            "duration_ms": 235534,
            "album_name": "Endless Smile"
        },
        {
            "pos": 2,
            "artist_name": "Degiheugi",
            "track_uri": "spotify:track:1vaffTCJxkyqeJY7zF9a55",
            "artist_uri": "spotify:artist:3V2paBXEoZIAhfZRJmo2jL",
            "track_name": "Some Beat in My Head",
            "album_uri": "spotify:album:2KrRMJ9z7Xjoz1Az4O6UML",
            "duration_ms": 268050,
            "album_name": "Dancing Chords and Fireflies"
        }, ...
    ],
}
\end{lstlisting}
% \end{minipage}

%\balancecolumns

\end{document}